\documentclass{article}
\usepackage{ijcai26}

\usepackage{times}
\usepackage{soul}
\usepackage{url}
\usepackage[hidelinks]{hyperref}
\usepackage[utf8]{inputenc}
\usepackage[small]{caption}
\usepackage{graphicx}
\usepackage{amsmath}
\usepackage{amsthm}
\usepackage{booktabs}
\usepackage{algorithm}
\usepackage{algorithmic}
\usepackage[switch]{lineno}

\usepackage{tabularx}
\usepackage{multirow}
\usepackage{array}
\usepackage{lscape}
\usepackage{xcolor}
\usepackage{graphicx}
\usepackage{booktabs,array,enumitem}
\usepackage{float}

\usepackage{ragged2e}
\usepackage[utf8]{inputenc}
\usepackage{longtable}
\usepackage{lipsum}
\usepackage{lineno}

\usepackage{caption}
\usepackage{subcaption}
\usepackage{float}

\definecolor{macaronpink}{RGB}{255,228,225}     
\definecolor{macaronblue}{RGB}{230,244,244}    
\definecolor{macarongreen}{RGB}{229,253,229}
\definecolor{macaronyellow}{RGB}{255,255,240}    
\definecolor{macaronpurple}{RGB}{243,230,243}    
\definecolor{macaronorange}{RGB}{255,245,238}    

\title{Human Tool: An MCP-Style Framework for Human-Agent Collaboration}

\author{
Yuanrong Tang$^1$\and
Huiling Peng$^2$\and
Bingxi Zhao$^3$\and
Hengyang Ding$^4$\and
Hanchao Song$^1$\and
Tianhong Wang$^{5,1}$\and
Chen Zhong$^1$\and
Jiangtao Gong$^{1,*}$\\
\affiliations
$^1$Tsinghua University\\
$^2$Zhejiang University\\
$^3$Beijing Jiaotong University\\
$^4$The Hong Kong University of Science and Technology\\
$^5$Anhui University\\
\emails
tangxtong2022@gmail.com,
gongjiangtao2@gmail.com
}

\begin{document}

\maketitle

\begin{abstract}

Human-AI collaboration faces growing challenges as AI systems increasingly outperform humans on complex tasks, while humans remain responsible for orchestration, validation, and decision oversight. To address this imbalance, we introduce Human Tool
, an MCP-style interface abstraction, building on recent Model Context Protocol designs, that exposes humans as callable tools within AI-led, proactive workflows. Here, “tool” denotes a coordination abstraction, not a reduction of human authority or responsibility.
Building on LLM-based agent architectures, we operationalize Human Tool by modeling human contributions through structured tool schemas of capabilities, information, and authority.
These schemas enable agents to dynamically invoke human input based on relative strengths and reintegrate it through efficient, natural interaction protocols.
We validate the framework through controlled studies in both decision-making and creative tasks, demonstrating improved task performance, reduced human workload, and more balanced collaboration dynamics compared to baseline systems. Finally, we discuss implications for human-centered AI design, highlighting how MCP-style human tools enable strong AI leadership while amplifying uniquely human strengths.

\end{abstract}

\section{Introduction}

Despite rapid advances in AI capabilities, effective human-AI collaboration remains elusive. A recent meta-analysis by Vaccaro et al. reveals a counterintuitive pattern: when humans outperform AI individually, collaboration tends to improve outcomes, whereas when AI outperforms humans, collaboration often degrades performance \cite{vaccaro2024combinations}. This meta-analytic pattern suggests that performance degradation in AI-advantaged settings may stem not from AI capability alone, but from how collaboration is commonly organized. In many systems, humans often retain responsibility for orchestration, validation, and coordination, even when AI can outperform them or scale better.

Prior work indicates that such human-centered orchestration can introduce systematic bottlenecks. Human limitations in attention, memory, consistency, and large-scale coordination become salient when AI is capable of handling complex workflows, information integration, and rule execution at scale \cite{Jarrahi2023}. In AI-advantaged tasks, assigning humans primary responsibility for coordination can introduce latency and variability, potentially undermining collaboration quality. This mechanism aligns with the degradation observed in AI-advantaged collaborations reported by Vaccaro et al. \cite{vaccaro2024combinations}, calling into question human-led coordination as a default design choice.


This tension has been widely discussed in HCI research on autonomy and division of labor, most notably in foundational work on mixed-initiative interaction, which formalizes how control and initiative dynamically shift between humans and automated systems based on context, uncertainty, and expected utility \cite{horvitz1999principles}. Subsequent work has extended this perspective to examine trust, shared understanding, and locus of control in mixed-initiative systems \cite{chiou2021trust_locus_control}, alongside research on levels of autonomy \cite{beer2014levels_robot_autonomy} and human-in-the-loop supervision \cite{sidouri2021jisa}.

\begin{figure}
\centering
\includegraphics[width=0.5\textwidth]{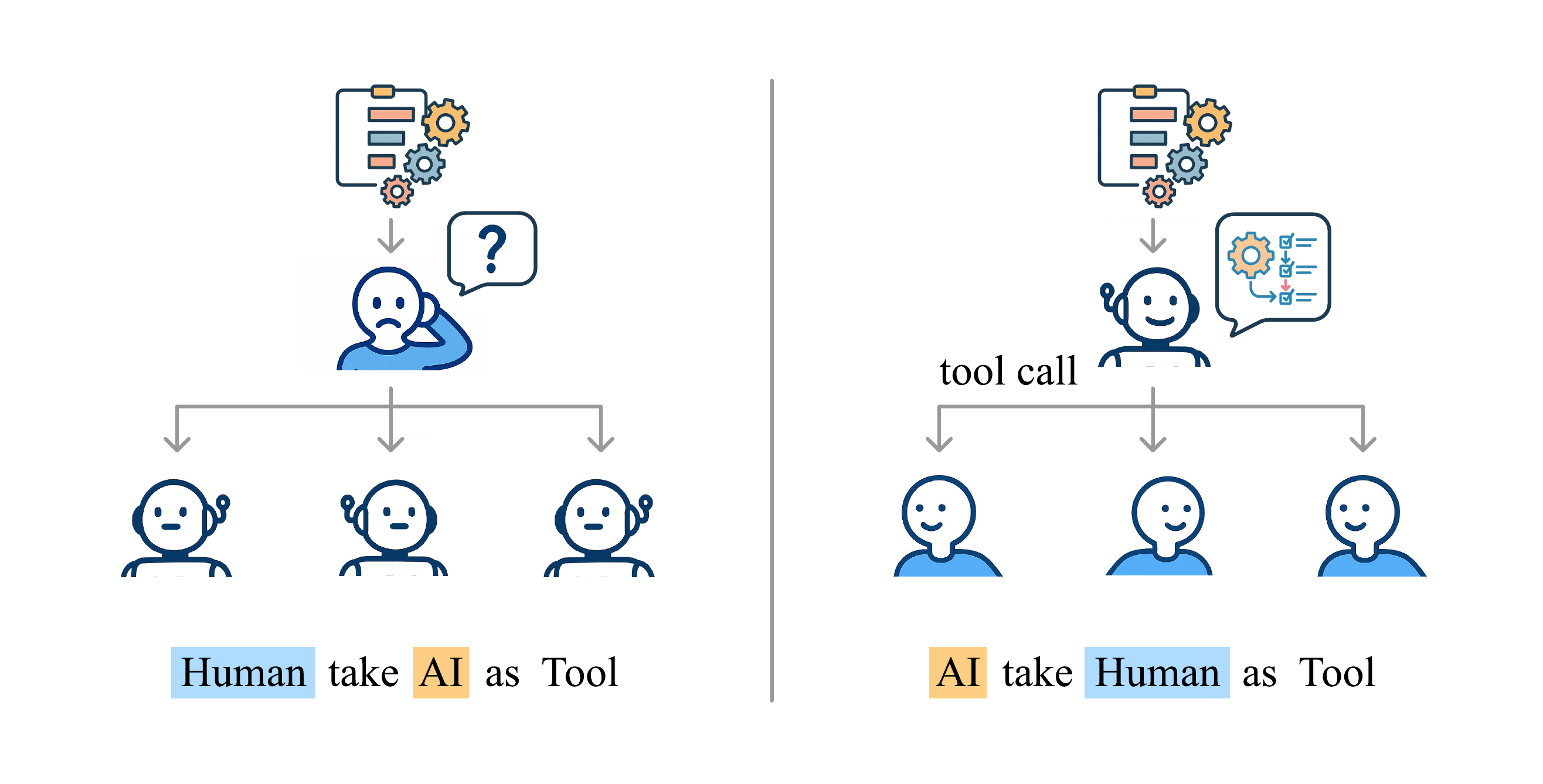}
\caption{Contrasting paradigms of human-AI collaboration: Human Tool versus AI Tool. In the Human Tool paradigm, humans are exposed to the agent as MCP-style callable interfaces rather than workflow leaders.}
\label{fig:teaser}
\end{figure}


We argue that this human-led design default is increasingly misaligned with contemporary AI capabilities. When AI can independently perform substantial portions of a task, the central design question should extend beyond how humans supervise AI to also include when and how AI should invoke human expertise. This reframes collaboration from a human-centered control problem to an AI-centered coordination problem, where AI actively manages workflows and recruits human input through structured, tool-like calls rather than continuous oversight.



To this end, we propose \textbf{Human Tool}, an AI-led and proactive collaboration paradigm in which AI takes ownership of workflow orchestration and treats humans as \emph{callable tools}. This perspective resonates with earlier work in human computation and crowd-powered systems, which conceptualized human contributors as on-demand services embedded within automated workflows (e.g., \cite{bigham2010vizwiz}). Building on recent advances in agent tool use and multi-component platforms \cite{schick2023toolformer}, Human Tool reframes this idea under modern LLM-based orchestration, modeling humans as \textbf{MCP-style callable interfaces} that can be selectively invoked when human judgment, creativity, or contextual reasoning provides comparative advantage.

Human Tool applies both to tasks AI could complete independently, where structured human invocation can improve robustness without reintroducing coordination bottlenecks, and to domains where AI cannot operate alone, where targeted human involvement is essential. In this sense, Human Tool provides a reusable framework for transforming AI-centric workflows into principled human-AI collaborations.

We operationalize Human Tool as an MCP-style abstraction through mechanisms for modeling human capabilities, deciding when to invoke human input, and enabling efficient AI-human communication. 
We evaluate this framework through controlled experiments with 32 participants across two task types motivated by conditions discussed in Vaccaro et al. \cite{vaccaro2024combinations}: a decision and optimization task where AI typically excels, and a creative task where human input remains critical. Compared to an AI Tool baseline, which follows the standard paradigm where humans lead task decomposition and invoke AI tools as needed, Human Tool improves performance, reduces perceived workload in AI-advantaged settings, and fosters more engaged collaboration dynamics.

\section{Related Work}

\subsection{Current Human-AI Collaboration Paradigms and Their Limitations}
\label{sec:RW1}

Prior work on human-AI systems ranges from user-invoked model assistance \cite{kacena2024use} and proactive recommendation interfaces \cite{Wen2025} to human oversight in autonomous and high-stakes settings \cite{heart2022intelligence}. Despite differences in initiative and autonomy, many approaches continue to place workflow orchestration and evaluation on humans, requiring users to decide when to involve AI and how to integrate its outputs.
This allocation can underutilize AI strengths while increasing coordination overhead. This role allocation is consistent with Vaccaro et al.’s meta-analysis showing that human-AI collaboration can degrade performance when AI outperforms humans individually \cite{vaccaro2024combinations}. Similar challenges have long been observed in crowdsourcing and human computation, where scaling human contribution depends critically on task decomposition, workflow control, and quality assurance \cite{quinn2011humancomputation}. Together, these findings motivate shifting coordination responsibility from humans to AI, while invoking human expertise selectively where it is most needed.

\subsection{Repositioning Roles: From Managerial Challenges to AI-Led Orchestration}
\label{sec:RW2}

Prior work shows that management and orchestration require task decomposition, role allocation, monitoring, and adaptation under interdependence, functions that are cognitively demanding and prone to failure due to limits in human attention, communication, and consistency \cite{Salimzadeh2024When,tankelevitch2024the}. By contrast, AI systems can scale information processing, maintain persistent tracking, and enforce consistency across workflows \cite{Raza2025}, making them well suited for orchestration in AI-advantaged settings \cite{vaccaro2024combinations}. At the same time, delegating managerial authority to AI raises concerns around trust calibration, accountability, and human agency \cite{Joseph2025Organization}, motivating mechanisms that enable AI to learn and exercise orchestration while selectively incorporating human judgment where it remains necessary.

\subsection{Integrating MCP-Style Tool-Calling Mechanisms into Human-AI Collaboration}
\label{sec:RW3}

Prior work on LLM tool use demonstrates that AI can decompose tasks and orchestrate executable workflows, from basic tool invocation \cite{schick2023toolformer} to standardized protocols supporting coordination across heterogeneous tools \cite{hou2025model}. Existing approaches, including proxy frameworks \cite{liu-etal-2025-llms} and personalization techniques \cite{Tan2023UserMI}, typically position humans as external validators or information sources. Earlier human computation and crowd-powered systems explicitly treated humans as callable services within computational pipelines (e.g., \cite{lasecki2011legion}), but typically relied on fixed task structures or external controllers rather than AI-driven, dynamic orchestration. 
In contrast, we model humans as first-class callable tools within the AI’s orchestration layer, analogous to computational resources \cite{zhou2024language} but subject to non-determinism, variable latency, refusal, negotiation, and cognitive constraints \cite{Feng2025}. Our Human Tool framework addresses these properties through explicit capability and consent boundaries, comparative allocation, and structured communication, enabling AI-led proactive orchestration while mitigating the human evaluation bottleneck identified by Vaccaro et al. \cite{vaccaro2024combinations}.

\section{Human Tool Framework}
\label{sec:Human tool framework}

\begin{figure*}[t]
\centering
\includegraphics[width=\textwidth]{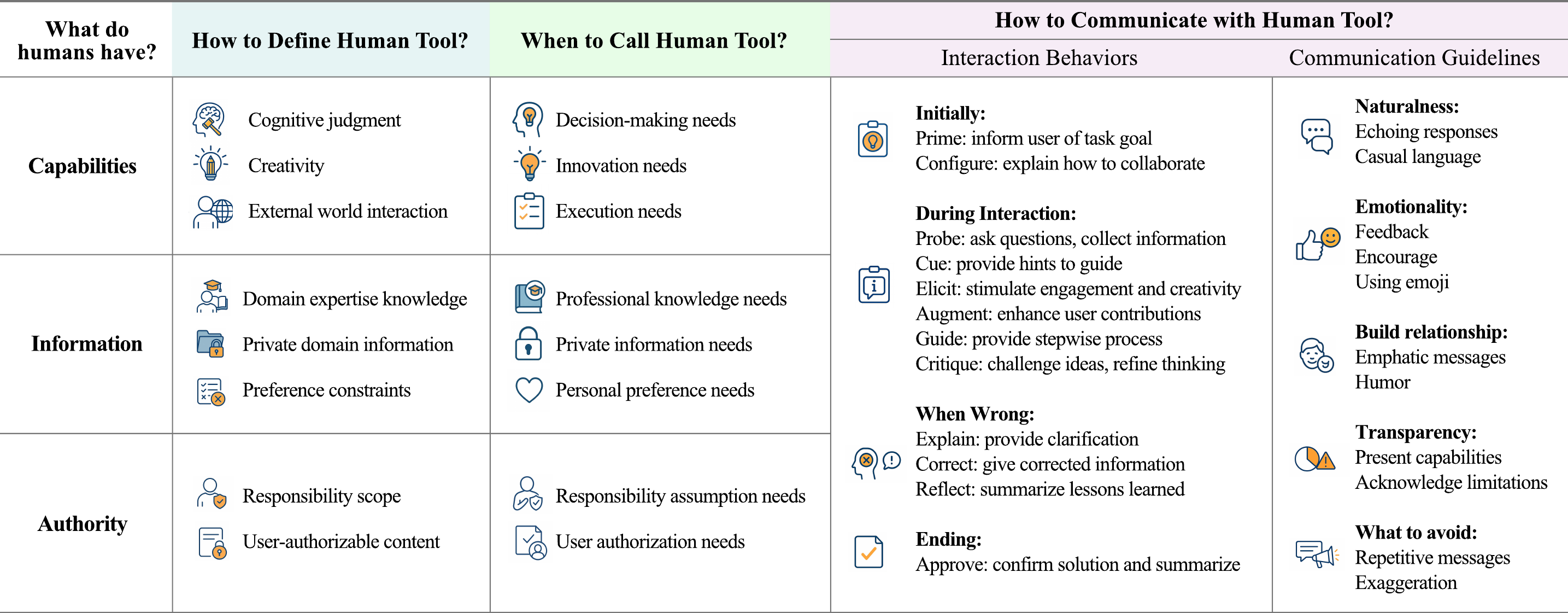}
\caption{An integrated framework for representing Human Tool within AI-managed workflows. It organizes three dimensions: How to define Human Tool, when to call them, and how to communicate effectively.}
\label{fig:human_tool_framework}
\end{figure*}


We operationalize Human Tool as an \textbf{MCP-style abstraction} that exposes human contributions to AI agents as callable tools within AI-led workflows. Rather than positioning humans as holistic evaluators or coordinators, this framework enables agents to invoke targeted human input when it provides comparative advantage. The framework is organized around three core questions aligned with tool-based orchestration: (1) how to define Human Tool through structured schemas of capabilities, information, and authority; (2) when to invoke them via dynamic allocation and safeguards; and (3) how to communicate with them using structured interaction behaviors that minimize coordination overhead. These components are formalized in Fig.~\ref{fig:human_tool_framework} and detailed in the following sections.

\subsection{\colorbox{macaronblue}{How to Define Human Tool?}}
\label{sec:How to define human tool?}

This section defines the Human Tool and how LLMs represent and invoke it.

\textbf{Definition.} A Human Tool is a structured, callable abstraction that represents a human contributor that enables reliable and efficient human-AI collaboration. It allows LLMs to decide when and how to involve humans by modeling them along three dimensions: Capabilities, Information, and Authority.

\begin{itemize}
    \item \textbf{Capabilities} capture forms of human contribution that remain difficult for AI, including cognitive judgment, creativity, and interaction with the external world \cite{vaccaro2024combinations,kahneman2011thinking}.
    
    \item \textbf{Information} represents knowledge resources held by humans, including domain expertise, private or situational information, and preference constraints that shape acceptable solutions \cite{tambe2025reskilling,farber2025peer}.
    
    \item \textbf{Authority} specifies decision boundaries and permissions, defining which decisions require human approval and what information may be shared with AI to ensure accountability and trustworthy collaboration \cite{siebert2021meaningfulhumancontrol}.
\end{itemize} 

Together, these dimensions standardize human contributions into a form that LLMs can reason about and invoke.

\textbf{LLM Interpretation.} Human Tools are instantiated as structured user profiles provided to the LLM via in-context prompts. In our system, participants self-assessed along these dimensions through a pre-study questionnaire (Appendix \ref{sec:Self-evaluation Scale}), and the resulting profiles were stored for invocation. The prompt format is shown in Appendix \ref{sec:prompt}, Fig.\ref{fig:prompt-understanding_human}.

\subsection{\colorbox{macarongreen}{When to Call Human Tool?}}
\label{sec:When to call human tool?}

This section identifies decision points in task decomposition where human involvement is required.

\textbf{Task decomposition.} Following hierarchical task analysis \cite{stanton2006hierarchical}, complex tasks are recursively decomposed into executable subtasks until reaching leaf nodes, enabling the system to reason about allocation decisions at each level \cite{vaccaro2024combinations}.

\textbf{Allocation decisions.} For each task node, the system decides whether AI or human should act based on relative strengths, drawing on interaction taxonomies \cite{shneiderman2020human}. Human involvement is invoked under three conditions:

\begin{itemize}

    \item \textbf{Capability complementarity (Capabilities):} tasks requiring novel reasoning, creativity, complex judgment, or physical interaction with the external world \cite{bansal2021does}.
    
    \item \textbf{Information exchange (Information):} tasks involving domain expertise, private constraints, or personal preferences that are inaccessible to AI \cite{lubars2019ask,ashktorab2021effects}.
    
    \item \textbf{Authority control (Authority):} decisions that require explicit human responsibility or authorization due to safety, privacy, or ethical considerations \cite{amershi2019guidelines}.

\end{itemize}

When these conditions are met, the system invokes human input; the corresponding prompt format is shown in Appendix \ref{sec:prompt}, Fig.~\ref{fig:prompt-task_decomposition}.

\subsection{\colorbox{macaronpurple}{How to Communicate with Human Tool?}}
\label{sec:How to communicate with human tool?}

Building on task decomposition, we address how AI communicates with humans during orchestration. Because humans are non-deterministic, variable in latency, and subject to refusal, negotiation, and cognitive limits \cite{Feng2025}, AI-led collaboration requires structured communication that elicits targeted input and reintegrates it without shifting holistic evaluation back to users \cite{wickramasinghe2020trustworthy}. We operationalize this along two dimensions: \textbf{what} to communicate (Interaction Behaviors) and \textbf{how} to communicate (Communication Guidelines), with prompts illustrated in Appendix \ref{sec:prompt}, Fig.~\ref{fig:prompt-Interaction_Behaviors}.

\textbf{Interaction behaviors.}
Following temporal interaction guidelines \cite{amershi2019guidelines} and agent communication primitives \cite{fipa2002acl}, we structure AI-human interaction into four stages aligned with why and when human input is needed:

\begin{itemize}
    \item \textbf{Initial:} establish context, goals, and collaboration boundaries.
    \item \textbf{During:} elicit preferences, constraints, judgments, or creative input through probing, cueing, guidance, or critique.
    \item \textbf{Error handling:} explain misunderstandings, correct errors, and repair shared state.
    \item \textbf{Ending:} confirm outcomes and close the loop.
\end{itemize}

\textbf{Communication guidelines.}
Adapting dialogue design principles for AI-led settings \cite{silva2024towards}, we adopt five lightweight guidelines to sustain engagement while minimizing user burden:
    \textbf{Naturalness:} maintain conversational flow with minimal prompting effort.
    \textbf{Emotionality:} provide calibrated encouragement and feedback \cite{bilquise2022emotionally}.
    \textbf{Relationship building:} foster a partner-like collaborative tone.
    \textbf{Transparency:} clearly communicate capabilities, limitations, and decision boundaries \cite{liu2022trustworthy}.
    \textbf{Avoidance:} minimize repetition and exaggeration to reduce fatigue.

\subsection{Implementing Human Tool as an MCP-style Interface}
\label{sec:tech_detail}

We implement the Human Tool framework as an \textbf{MCP-style callable interface}, where human contributions are exposed to the agent as standardized, invocable resources rather than as workflow leaders. In this abstraction, each human is modeled as a tool with explicit input-output boundaries, invocation conditions, and interaction protocols, enabling the agent to orchestrate human participation in the same manner as other tools.

The system is implemented with a Python backend using LangGraph to support graph-based orchestration, with GPT-4o as the reasoning core. User profiles and interaction history are stored in MySQL, and a React TypeScript frontend communicates via FastAPI, enabling structured orchestration with real-time LLM reasoning.

\begin{figure}[!htbp]
  \centering
  \includegraphics[width=1\linewidth]{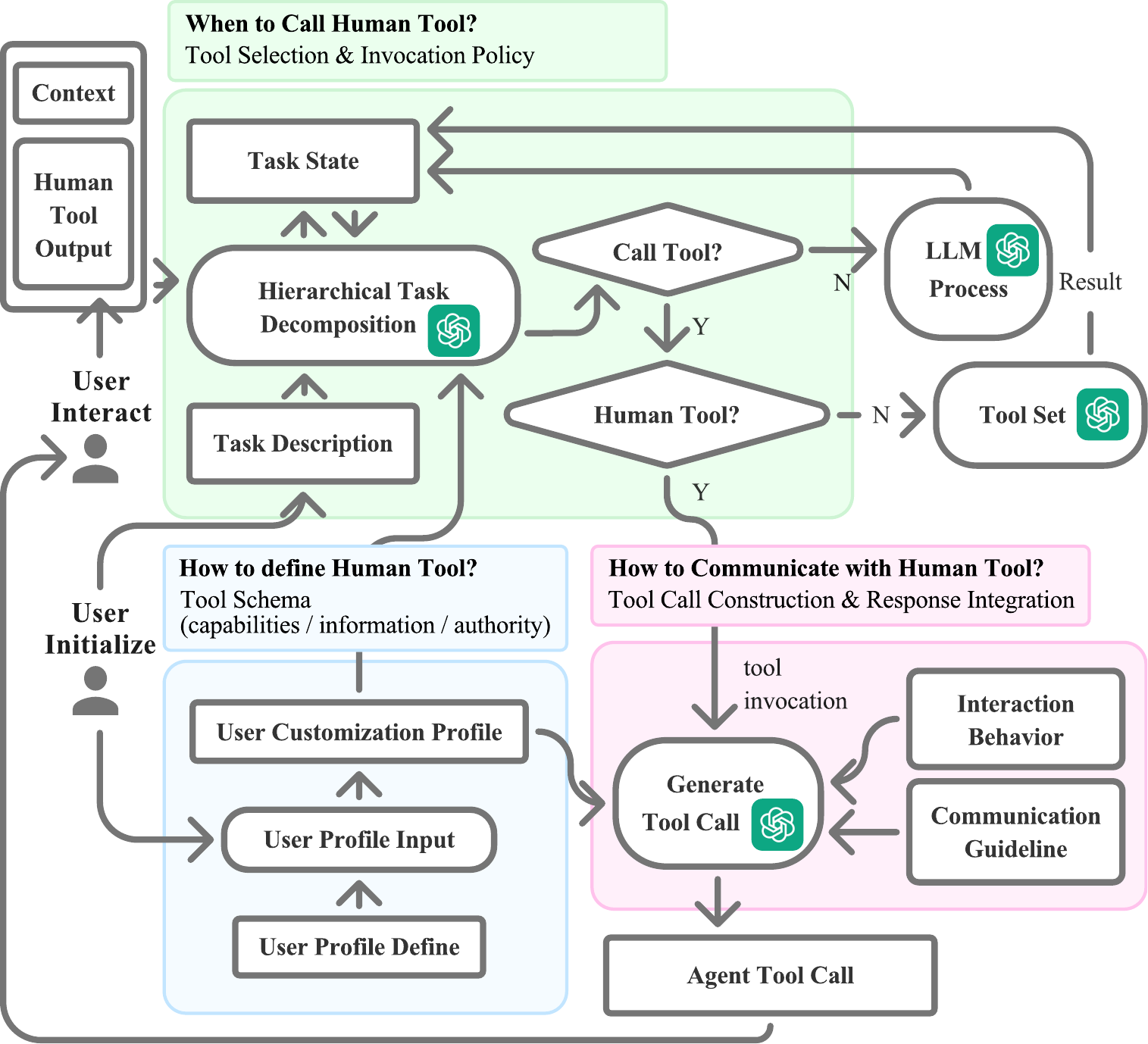}
  \caption{MCP-style orchestration architecture of the Human Tool framework, illustrating tool schema definition, invocation decisions, and tool call-response integration.}
  \label{fig:tech}
\end{figure}

As shown in Fig.~\ref{fig:tech}, execution follows three stages that correspond directly to the MCP abstraction. First, \colorbox{macaronblue}{How to Define Human Tool?} specifies the Human Tool’s interface by encoding capabilities, information access, and authority boundaries. Second, \colorbox{macarongreen}{When to Call Human Tool?} determines invocation points during hierarchical task decomposition, analogous to tool-selection logic in MCP-style systems. When invocation conditions are met, the workflow enters \colorbox{macaronpurple}{How to Communicate with Human Tool?}, where the agent performs a structured “tool call” by generating guideline-compliant prompts, receiving human responses, and updating task state.

Through this design, human input is integrated via explicit calls rather than continuous supervision, allowing the LLM to function as both the orchestrator and the caller of Human Tools while maintaining controllability and modularity.

\section{Experiments}

This section describes the study design, including tasks, participants, procedures, and measures. We investigate whether Human Tool improves collaborative performance and reduces workload (RQ1), generalizes across task types (RQ2), and enhances user engagement and collaborative human-AI interaction (RQ3).

\subsection{Overview}
We conducted a controlled user study comparing \textbf{Human Tool} against an \textbf{AI Tool baseline}. Both conditions used the same backend models (GPT-4o with identical decoding settings), interface, memory and context handling, tool routing policies, logging, task-matched domain tools, and identical time and turn budgets. The only manipulation was whether the system could invoke \emph{human-as-a-tool} (enabled in Human Tool, disabled in AI Tool). The baseline reflects the standard “chat with an LLM plus fixed domain tools” paradigm commonly adopted in current agent systems. All procedures were approved by an independent IRB, and the study was run online via VooV Meeting.

\subsection{Participants}
We recruited 33 participants and excluded one due to system errors, yielding a final sample of 32. Participants were aged 21-35 (\emph{M}=25.24, \emph{SD}=3.38). They were assigned to one task domain (Travel Planning or Story Writing), with 16 participants per domain. Demographics and prior LLM experience were balanced across domains; full details are reported in Appendix (Table~\ref{tab:participant_demographics_ab}). Participants received a stipend of 150 CNY (approximately USD 21).

\subsection{Procedure}
We used a within-subject, counterbalanced design. Within each domain, every participant completed two comparable task instances, one per condition, with both the condition order (AB/BA) and the mapping between task instances and conditions fully counterbalanced to mitigate order and content effects. One day before the study, participants completed a brief tutorial using a trial system (same frontend, different backend) and filled out a Human Tool configuration questionnaire (Appendix~\ref{sec:Self-evaluation Scale}). On the study day, each run lasted 15 minutes under fixed time and turn limits, followed by task-specific questionnaires and a short semi-structured interview. After both runs, participants completed a final comparative interview; total session time was approximately 1.5 hours.

\textbf{Tasks.} We evaluated two task types that represent distinct collaboration modes \cite{vaccaro2024combinations}. (1) \emph{Travel Planning} is a constrained decision-making task built on the TravelPlanner system and dataset \cite{10.5555/3692070.3694316}. Participants co-developed a multi-day itinerary under budget and dataset-grounded constraints (e.g., valid flight numbers, lodging rules, and non-repeating dining), using scenario prompts such as planning a Las Vegas to Santa Maria trip or an Ithaca to Newark trip with preference constraints. In this task, the agent treats the participant as a Human Tool and issues MCP-style calls at decision-critical nodes during itinerary construction. Specifically, Human Tool calls are triggered for preference clarification, responsibility assignment (e.g., trade-offs between cost and comfort), and constraint resolution when multiple valid branches exist.
(2) \emph{Story Writing} is an open-ended creative task adapted from narrative prompt designs \cite{10.1145/3491102.3502030}. Participants collaborated with the system to produce a complete story outline from scenario premises (e.g., a post-apocalyptic world with rediscovered magic, or an anti-aging drug that creates intergenerational conflict), emphasizing both content qualities (character depth, authenticity, theme clarity) and structural qualities (causal coherence, progressive tension, turning points, multidimensional conflict). 
In Story Writing, the Human Tool functions as a non-deterministic MCP-style resource that is actively orchestrated by the agent throughout the creative process. The agent maintains end-to-end control over narrative structure and progression, while repeatedly issuing Human Tool calls to elicit creative ideation, critique, and aesthetic judgment. These calls occur at both high-level planning stages (e.g., concept selection and plot structuring) and during content generation and refinement, with human outputs integrated as tool responses that guide subsequent narrative planning and control.
Full task materials appear in Appendix (Table~\ref{tab:task_detail}).

\subsection{Metrics}
\textbf{Quantitative measures.} After each run, participants completed standardized questionnaires covering usability and experience, including SUS \cite{lewis2018system}, USE-SF \cite{o2018practical}, Rating Scale Mental Effort \cite{zijlstra1985construction}, RCS \cite{burgoon1987relational}, and ASCC \cite{mavri2020assessment}, plus a custom scale targeting system-specific features (Appendix~\ref{sec:Self-Design Scale}). For story writing, we additionally collected CSI \cite{10.1145/2617588}. We complemented self-reports with outcome measures: for Travel Planning, scripts evaluated constraint compliance against the dataset \cite{10.5555/3692070.3694316} (per-constraint pass rates and an overall pass); for Story Writing, we collected pairwise preferences from 20 independent raters and derived head-to-head win rates. We also logged human-tool invocations to quantify when, why, and how human input was requested and integrated.

\textbf{Qualitative measures.} We conducted brief semi-structured interviews after each condition and a final comparative interview after both conditions (Appendix~\ref{sec:Semi-structured Interviews Question}). Interviews were recorded, transcribed (ASR followed by manual correction), and analyzed using reflexive thematic analysis \cite{braun2006using} with back-to-back coding to support reliability.

\section{Results}
\label{sec:result}

Our findings show that the Human Tool Group improves human--AI collaboration in three AI-relevant respects: task performance and efficiency (\textbf{RQ1}, Sec.\ref{sec:RQ1}), adaptive human-input elicitation that supports decision-making (\textbf{RQ2}, Sec.\ref{sec:RQ2}), and reduced coordination burden with better time regulation in collaboration (\textbf{RQ3}, Sec.\ref{sec:RQ3}). We report only significant results and prioritize performance-, time-, and workload-facing metrics, complemented by brief qualitative evidence.

\subsection{\textbf{Improved task performance and reduced workload through targeted allocation and scaffolding (RQ1)}}
\label{sec:RQ1}

\begin{figure}[!htbp]
  \centering
  \includegraphics[width=0.6\linewidth]{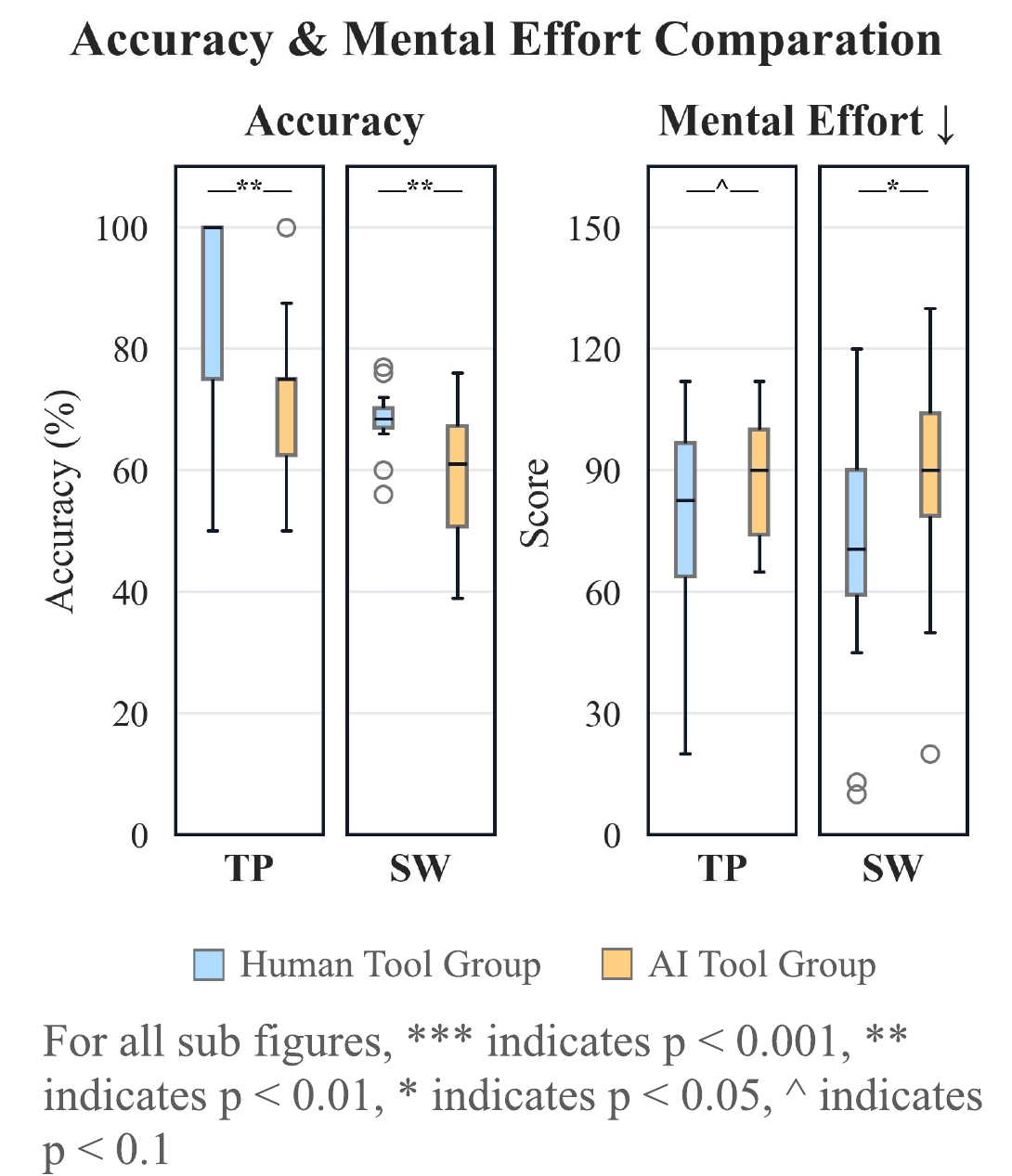}  
  \caption{Comparison of Accuracy and Mental Effort Across Tasks (Travel Planning vs. Story Writing) Between Human Tool Group (blue) and AI Tool Group (orange)}
  \label{fig:acc-effort-tasks-comparison}
\end{figure}

We evaluated the Human Tool framework using objective task performance and subjective cognitive load, complemented by interviews. Figure~\ref{fig:acc-effort-tasks-comparison} summarizes results for Travel Planning (TP) and Story Writing (SW).

\textbf{Task accuracy and quality.} Human Tool Group outperformed AI Tool Group on both tasks. In TP, accuracy (percent correct) was higher for Human Tool Group (\emph{M}=86.72, \emph{SD}=18.52) than AI Tool Group (\emph{M}=72.66, \emph{SD}=12.26), a 19.34\% gain (paired-samples \emph{t}=3.58, \emph{p}=.003 $<$ .01). In SW (LLM auto-scored), Human Tool Group also exceeded AI Tool Group (\emph{M}=68.38, \emph{SD}=5.11 vs.\ \emph{M}=58.56, \emph{SD}=11.05), a 14.35\% improvement (\emph{t}=3.22, \emph{p}=.006 $<$ .01). Human ratings corroborated this advantage: SW win-rate favored Human Tool Group (\emph{M}=0.611, \emph{SD}=0.194) over AI Tool Group (\emph{M}=0.371, \emph{SD}=0.211), \emph{t}=3.538, \emph{p}=.003 $<$ .01. Participants attributed these gains to clearer decomposition and stepwise scaffolding (TP-9: \textit{“the step-by-step guidance made me \textbf{more confident in the final plans}.”}).

\textbf{Collaboration support.} Consistent with improved task outcomes, overall CSI was higher for Human Tool (\emph{M}=75.48, \emph{SD}=10.22) than AI Tool (\emph{M}=52.83, \emph{SD}=15.58), \emph{t}=6.92, \emph{p}=$4.90\times10^{-6}$ $<$ .001 (See Appendix \ref{sec:Detailed Experimental Results}, Fig.~\ref{fig:csi-groups-comparison} for details).

\textbf{Cognitive load and effort (efficiency proxy).} Human Tool Group reduced mental effort in SW: \emph{M}=70.625 (\emph{SD}=31.07) vs.\ \emph{M}=87.875 (\emph{SD}=27.14), a 19.63\% decrease (\emph{t}=-2.73, \emph{p}=.016 $<$ .05). TP showed a similar direction but did not reach significance (\emph{t}=-1.88, \emph{p}=.080, n.s.). Interviews suggest the baseline imposed higher “management burden” (TP-11); users described needing repeated prompt crafting and checking, whereas Human Tool Group \textit{“\textbf{did the dirty work}”} (TP-2), shifting effort from coordination toward substantive decision-making (SW-2: \textit{“less time managing… \textbf{focus on higher-value reasoning}.”}).

Across planning and creative tasks, Human Tool Group improved objective performance while reducing cognitive effort, consistent with targeted allocation and scaffolding that offload coordination and elevate high-value human input.

\subsection{\textbf{Task-type adaptivity and high usability via appropriate what/when invocation of Human Tool (RQ2)}}
\label{sec:RQ2}

We analyzed task-type adaptivity in \emph{what} human inputs were requested, \emph{when} they were requested, and \emph{how} the system communicated, using activation logs and post-task ratings (Fig.~\ref{fig:data-count-TP}, Fig.~\ref{fig:cross-task-system-behavior-quality}; items were aggregated into grouped dimensions following Table~\ref{fig:human_tool_framework}). Overall usability was higher for the Human Tool Group (SUS): TP \emph{M}=70.89 (\emph{SD}=10.96) vs.\ 58.22 (\emph{SD}=18.98), \emph{t}=2.81, \emph{p}=.014; SW \emph{M}=79.79 (\emph{SD}=7.93) vs.\ 60.21 (\emph{SD}=15.95), \emph{t}=5.20, \emph{p}=$1.08\times10^{-4}$.

\textbf{What to request (decision-relevant human inputs).} Compared to AI Tool Group, Human Tool Group more strongly elicited and reused user preferences and responsibility boundaries: using personal information/preferences (\emph{M}=5.56 vs.\ 3.63, \emph{t}=3.91, \emph{p}=.0014) and clarifying responsibility/permissions (\emph{M}=5.69 vs.\ 3.69, \emph{t}=4.30, \emph{p}=$6.40\times10^{-4}$). In TP, leveraging user knowledge/preferences was also higher (\emph{M}=6.13 vs.\ 3.31, \emph{t}=7.03, \emph{p}=$4.09\times10^{-6}$).

\textbf{When to request (timing at key decision points).} Logs show invocations concentrated at decision-critical moments (e.g., preference clarification and choice points) rather than low-value confirmations. Self-reports match this pattern: TP Seek Help (\emph{M}=5.25 vs.\ 3.25, \emph{t}=5.31, \emph{p}=$8.68\times10^{-5}$) and Identify Knowledge (\emph{M}=5.93 vs.\ 3.80, \emph{t}=4.67, \emph{p}=$3.00\times10^{-4}$) were higher; SW showed the same gains (Seek Help: \emph{p}=$4.25\times10^{-4}$; Identify Knowledge: \emph{p}=$7.21\times10^{-4}$) and higher Request Authorization (\emph{M}=5.63 vs.\ 4.25, \emph{t}=3.15, \emph{p}=.0066). Participants summarized this as asking “at the right time” after options were surfaced (TP-2).

\textbf{How to communicate (scaffolding signals).} Adaptive prompting amplified core scaffolding behaviors that directly support task progress. In TP, Guide (\emph{p}=$8.97\times10^{-4}$), Elicit (\emph{p}=$2.62\times10^{-4}$), and Decision Augment (\emph{p}=$6.47\times10^{-4}$) were higher; in SW, Cue (\emph{p}=$1.03\times10^{-5}$), Guide (\emph{p}=$3.88\times10^{-5}$), and Probe (\emph{p}=$3.42\times10^{-4}$) increased. Users emphasized the resulting stepwise clarity: \textit{“Almost \textbf{no moments of confusion}; it walked me through in order.”} (TP-8).

\subsection{\textbf{Reduced coordination burden and improved time regulation in collaboration (RQ3)}}
\label{sec:RQ3}

We evaluated \textbf{RQ3} via engagement and collaboration outcomes in TP and SW (Fig.~\ref{fig:use-tasks-comparison}, Fig.~\ref{fig:relation-tasks-comparison}, Fig.~\ref{fig:collab-tasks-comparison}). We report measures most relevant to efficiency/time and collaboration quality, retaining significant effects.

\textbf{Efficiency and time-related experience.} On Perceived Usability (PU-S; lower is better), participants reported lower interaction burden with Human Tool Group in TP (composite: \emph{M}=1.96 vs.\ 2.81, \emph{t}=-3.40, \emph{p}=.0039) and SW (composite: \emph{M}=1.42 vs.\ 3.15, \emph{t}=-6.23, \emph{p}=$1.61\times10^{-5}$). For time perception, SW showed higher immersion on FA-S.2 (“time flew by”: \emph{M}=4.69 vs.\ 3.88, \emph{SD}=0.60/1.15, \emph{t}=3.57, \emph{p}=.0028). Consistent with improved pacing/coordination, Time Regulation and Achievement (ASCC-TRA) increased in SW (\emph{M}=5.09 vs.\ 4.16, \emph{SD}=1.34/1.21, \emph{t}=5.1551, \emph{p}=$1.17\times10^{-4}$).

\textbf{Collaboration dynamics.} On RCS, Human Tool Group supported more reciprocal coordination in TP, including stronger listening (\emph{M}=6.27 vs.\ 5.07, \emph{SD}=0.70/1.49, \emph{t}=3.15, \emph{p}=.0070) and reduced coldness (\emph{M}=2.20 vs.\ 3.73, \emph{SD}=0.77/1.91, \emph{t}=-3.44, \emph{p}=.0040) and distance (\emph{M}=2.40 vs.\ 3.87, \emph{SD}=1.06/1.77, \emph{t}=-3.29, \emph{p}=.0054). In SW, participants also reported more partner-like interaction, with higher deepening moves (\emph{M}=5.44 vs.\ 3.06, \emph{SD}=1.21/1.57, \emph{t}=5.44, \emph{p}=$6.84\times10^{-5}$) and reduced formality (\emph{M}=2.06 vs.\ 3.19, \emph{SD}=1.18/1.60, \emph{t}=-3.20, \emph{p}=.0060). On ASCC, Synergistic Social Collaboration (SSC) was higher in TP (\emph{M}=5.36 vs.\ 4.46, \emph{SD}=0.80/1.25, \emph{t}=2.66, \emph{p}=.0177) and SW (\emph{M}=5.71 vs.\ 4.09, \emph{SD}=0.83/1.29, \emph{t}=4.8037, \emph{p}=$2.32\times10^{-4}$).

Qualitatively, participants described Human Tool Group as reducing coordination overhead while enabling deeper joint exploration: \textit{“Human Tool Group felt like a \textbf{partner}; we exchanged opinions and sparked ideas… explore the task more deeply”} (TP-6), and \textit{“It truly \textbf{interacted with me as a friend}”} (SW-2).

\section{Discussion}

This study evaluates the \emph{Human Tool} paradigm, treating humans as callable resources within an AI-led workflow, across a planning task (TP) and a creative task (SW). Overall, results consistently favor the Human Tool Group across objective performance, time-/effort-related experience, and collaboration dynamics, while interviews suggest that the gains come from shifting orchestration work (decomposition, coordination, progress tracking) from humans to the agent and reserving human effort for high-value judgments.

\subsection{Summary of findings}

\textbf{Performance and output quality (RQ1).} The Human Tool Group produced higher-quality outcomes on both tasks. In TP, accuracy improved from \emph{M}=72.66 to 86.72 (paired \emph{t}=3.58, \emph{p}=.003). In SW, LLM auto-scored quality increased from \emph{M}=58.56 to 68.38 (paired \emph{t}=3.22, \emph{p}=.0057), and human win-rate favored the Human Tool Group (\emph{M}=0.611 vs.\ 0.371; paired \emph{t}=3.538, \emph{p}=.0030). Participants attributed these gains to clearer decomposition and stepwise scaffolding, e.g., increased confidence in the final plans (TP-9) and better retention of the user’s intended storyline while improving structure (SW-1).

\textbf{Time and effort (RQ1/RQ3).} Mental effort decreased in SW (\emph{M}=70.625 vs.\ 87.875; \emph{t}=-2.73, \emph{p}=.016), with interview evidence that the baseline condition imposed more “management burden” (TP-11). In engagement measures that map most directly onto time/burden, the Human Tool Group reduced perceived usability burden (PU-S; lower is better) in both TP (\emph{t}=-3.40, \emph{p}=.0039) and SW (\emph{t}=-6.23, \emph{p}=$1.61\times10^{-5}$), and increased time-flew-by immersion in SW (FA-S.2; \emph{t}=3.57, \emph{p}=.0028). These patterns align with a simple mechanism: the agent absorbs coordination overhead, so users spend proportionally more time on substantive decisions rather than prompt engineering and checking.

\textbf{Adaptive invocation and system usefulness (RQ2).} The Human Tool Group achieved higher overall usability (SUS) in both tasks: TP (\emph{t}=2.81, \emph{p}=.014) and SW (\emph{t}=5.20, \emph{p}=$1.08\times10^{-4}$). Logs and self-reports indicate that calls concentrated at decision points (preferences, choices, and knowledge elicitation) rather than low-value confirmations. Participants described this as “asking at the right time” (TP-2) and as a structured, low-confusion flow (TP-8).

\textbf{More partner-like collaboration (RQ3).} Beyond task success, the Human Tool Group supported more collaborative dynamics. Quantitatively, synergistic social collaboration (ASCC-SSC) improved in TP (\emph{t}=2.66, \emph{p}=.0177) and SW (\emph{t}=4.8037, \emph{p}=$2.32\times10^{-4}$), and time regulation/achievement increased in SW (\emph{t}=5.1551, \emph{p}=$1.17\times10^{-4}$). Qualitatively, participants repeatedly described the system as a “partner” rather than a one-way tool (TP-6) and emphasized smoother co-development of ideas in writing (SW-2).

\subsection{What the Human Tool reframing adds}

A practical implication of this reframing is that \emph{highly capable AI can improve collaboration by assuming responsibility for orchestration}. In our setting, the Human Tool Group performed better not because humans did more, but because they did less of the wrong kind of work. The agent managed the operational layer, planning workflows, sequencing steps, and tracking constraints, while humans were invoked for information that is costly or unreliable for the agent to infer, such as preferences, intent, boundaries, creative priors, and subjective judgment. This division of labor improved both \emph{performance} (higher-quality outputs) and \emph{experience} (lower effort and greater immersion).

This reframing also clarifies how a system can remain human-centered even when humans appear, superficially, to function as “tools” that are called upon when needed. Although human input is episodic rather than continuous, the system’s evaluation criteria remain grounded in human subjective experience: usefulness, satisfaction, authorship, and perceived quality. From this perspective, humans are not reduced to objects within the system, but remain the ultimate locus of value. What changes is not whose interests are served, but \emph{how} human agency is expressed.

Rather than centering humans through constant supervision, the Human Tool pattern centers humans by making their contributions more consequential. When invoked, humans provide decisions that meaningfully shape the trajectory of the plan or the narrative, and the system then carries those decisions forward with consistency and scale. Interview feedback suggests that this mode of collaboration was experienced less as being “managed” and more as being supported, a lightweight yet respectful interaction style that preserved users’ sense of authorship (e.g., retaining ideas while tightening structure in SW). At its best, the Human Tool pattern shifts interaction from iterative correction to joint progress.



\subsection{Limitations and future work}

This evaluation focuses on two tasks (TP, SW) in a controlled study setting, which may limit generalizability to longer-horizon or higher-stakes work. Future studies should test broader domains and incorporate more direct time-based measures (e.g., completion time, revision/backtracking) to better quantify efficiency. Finally, because some participants reported feeling “carried along” by strong guidance, future designs can add lightweight checkpoints to preserve proactive orchestration while making user control and authorship more explicit.

\section{Conclusion}


In conclusion, the Human Tool framework introduces an MCP-style abstraction for human-AI collaboration, in which human expertise is exposed as a callable interface within AI-led workflows. Across both decision-making and creative tasks, our results show that invoking humans selectively, rather than positioning them as continuous supervisors, improves performance while reducing cognitive workload. By allowing agents to orchestrate workflows and call on human judgment where it offers clear comparative advantage, Human Tool supports more efficient and balanced collaboration. More broadly, this work points to the potential of extending MCP-style abstractions beyond computational tools to human contributors as a useful design direction for human-centered AI systems.

\bibliographystyle{named}
\bibliography{./sample-base}

@String{Computing = "Computing" }

@String{Academic = "Academic Press" }

@String{Springer = "Springer-Verlag" }

@inproceedings{amershi2019guidelines,
  title={Guidelines for human-AI interaction},
  author={Amershi, Saleema and Weld, Dan and Vorvoreanu, Mihaela and Fourney, Adam and Nushi, Besmira and Collisson, Penny and Suh, Jina and Iqbal, Shamsi and Bennett, Paul N and Inkpen, Kori and others},
  booktitle={Proceedings of the 2019 chi conference on human factors in computing systems},
  pages={1--13},
  year={2019}
}

@article{silva2024towards,
  title={Towards user-centric guidelines for chatbot conversational design},
  author={Silva, Geovana Ramos Sousa and Canedo, Edna Dias},
  journal={International Journal of Human--Computer Interaction},
  volume={40},
  number={2},
  pages={98--120},
  year={2024},
  publisher={Taylor \& Francis}
}

@article{stanton2006hierarchical,
  title={Hierarchical task analysis: Developments, applications, and extensions},
  author={Stanton, Neville A},
  journal={Applied ergonomics},
  volume={37},
  number={1},
  pages={55--79},
  year={2006},
  publisher={Elsevier}
}

@article{shneiderman2020human,
  title={Human-centered artificial intelligence: Reliable, safe \& trustworthy},
  author={Shneiderman, Ben},
  journal={International Journal of Human--Computer Interaction},
  volume={36},
  number={6},
  pages={495--504},
  year={2020},
  publisher={Taylor \& Francis}
}

@article{lubars2019ask,
  title={Ask not what AI can do, but what AI should do: Towards a framework of task delegability},
  author={Lubars, Brian and Tan, Chenhao},
  journal={Advances in neural information processing systems},
  volume={32},
  year={2019}
}

@inproceedings{bansal2021does,
  title={Does the whole exceed its parts? the effect of ai explanations on complementary team performance},
  author={Bansal, Gagan and Wu, Tongshuang and Zhou, Joyce and Fok, Raymond and Nushi, Besmira and Kamar, Ece and Ribeiro, Marco Tulio and Weld, Daniel},
  booktitle={Proceedings of the 2021 CHI conference on human factors in computing systems},
  pages={1--16},
  year={2021}
}

@inproceedings{ashktorab2021effects,
  title={Effects of communication directionality and AI agent differences in human-AI interaction},
  author={Ashktorab, Zahra and Dugan, Casey and Johnson, James and Pan, Qian and Zhang, Wei and Kumaravel, Sadhana and Campbell, Murray},
  booktitle={Proceedings of the 2021 CHI conference on human factors in computing systems},
  pages={1--15},
  year={2021}
}

@inproceedings{wickramasinghe2020trustworthy,
  title={Trustworthy AI development guidelines for human system interaction},
  author={Wickramasinghe, Chathurika S and Marino, Daniel L and Grandio, Javier and Manic, Milos},
  booktitle={2020 13th International Conference on Human System Interaction (HSI)},
  pages={130--136},
  year={2020},
  organization={IEEE}
}

@article{vaccaro2024combinations,
  title={When combinations of humans and AI are useful: A systematic review and meta-analysis},
  author={Vaccaro, Michelle and Almaatouq, Abdullah and Malone, Thomas},
  journal={Nature Human Behaviour},
  volume={8},
  number={12},
  pages={2293--2303},
  year={2024},
  publisher={Nature Publishing Group UK London}
}

@misc{fipa2002acl,
  title={FIPA Communicative Act Library Specification},
  author={{Foundation for Intelligent Physical Agents (FIPA)}},
  howpublished={\url{http://www.fipa.org/specs/fipa00037/SC00037J.html}},
  year={2002}
}

@article{bilquise2022emotionally,
  title={Emotionally intelligent chatbots: A systematic literature review},
  author={Bilquise, Ghazala and Ibrahim, Samar and Shaalan, Khaled},
  journal={Human Behavior and Emerging Technologies},
  volume={2022},
  number={1},
  pages={9601630},
  year={2022},
  publisher={Wiley Online Library}
}

@article{liu2022trustworthy,
  title={Trustworthy ai: A computational perspective},
  author={Liu, Haochen and Wang, Yiqi and Fan, Wenqi and Liu, Xiaorui and Li, Yaxin and Jain, Shaili and Liu, Yunhao and Jain, Anil and Tang, Jiliang},
  journal={ACM Transactions on Intelligent Systems and Technology},
  volume={14},
  number={1},
  pages={1--59},
  year={2022},
  publisher={ACM New York, NY}
}

@book{kahneman2011thinking,
  title={Thinking, fast and slow},
  author={Kahneman, Daniel},
  publisher={Farrar, Straus and Giroux},
  year={2011}
}

@article{tambe2025reskilling,
  title={Reskilling the Workforce for AI: Domain Experts, Algorithms, and Organizational Performance},
  author={Tambe, Prasanna},
  journal={Management Science},
  volume={2},
  year={2025}
}

@article{farber2025peer,
  title={Comparing human and AI expertise in the academic peer review process},
  author={Farber, S.},
  journal={Learning \& Individual Differences},
  volume={xx},
  pages={yy--zz},
  year={2025},
  doi={10.1080/07294360.2024.2445575}
}

@article{siebert2021meaningfulhumancontrol,
  title={Meaningful human control: Actionable properties for AI system development},
  author={Siebert, Luciano Cavalcante and Lupetti, Maria Luce and Aizenberg, Evgeni and Beckers, Niek and Zgonnikov, Arkady and Veluwenkamp, Herman and Abbink, David and Giaccardi, Elisa and Houben, Geert-Jan and Jonker, Catholijn M. and van den Hoven, Jeroen and Forster, Deborah and Lagendijk, Reginald L.},
  journal={Proceedings of the AAAI/ACM Conference on AI, Ethics, and Society},
  year={2021},
  note={Formal publication},
  url={https://ojs.aaai.org/index.php/aaai/article/view/17691}
}

@inproceedings{10.1145/3491102.3502030,
author = {Lee, Mina and Liang, Percy and Yang, Qian},
title = {CoAuthor: Designing a Human-AI Collaborative Writing Dataset for Exploring Language Model Capabilities},
year = {2022},
isbn = {9781450391573},
publisher = {Association for Computing Machinery},
address = {New York, NY, USA},
url = {https://doi.org/10.1145/3491102.3502030},
doi = {10.1145/3491102.3502030},
booktitle = {Proceedings of the 2022 CHI Conference on Human Factors in Computing Systems},
articleno = {388},
numpages = {19},
location = {New Orleans, LA, USA},
series = {CHI '22}
}

@inproceedings{10.5555/3692070.3694316,
author = {Xie, Jian and Zhang, Kai and Chen, Jiangjie and Zhu, Tinghui and Lou, Renze and Tian, Yuandong and Xiao, Yanghua and Su, Yu},
title = {TravelPlanner: a benchmark for real-world planning with language agents},
year = {2024},
publisher = {JMLR.org},
booktitle = {Proceedings of the 41st International Conference on Machine Learning},
articleno = {2246},
numpages = {24},
location = {Vienna, Austria},
series = {ICML'24}
}

@article{lewis2018system,
  title={The system usability scale: past, present, and future},
  author={Lewis, James R},
  journal={International Journal of Human--Computer Interaction},
  volume={34},
  number={7},
  pages={577--590},
  year={2018},
  publisher={Taylor \& Francis}
}

@article{o2018practical,
  title={A practical approach to measuring user engagement with the refined user engagement scale (UES) and new UES short form},
  author={O’Brien, Heather L and Cairns, Paul and Hall, Mark},
  journal={International Journal of Human-Computer Studies},
  volume={112},
  pages={28--39},
  year={2018},
  publisher={Elsevier}
}

@book{zijlstra1985construction,
  title={The construction of a scale to measure perceived effort},
  author={Zijlstra, Ferdinand RH and Van Doorn, L},
  year={1985},
  publisher={University of Technology}
}

@article{burgoon1987relational,
  title={Relational communication, satisfaction, compliance-gaining strategies, and compliance in communication between physicians and patients},
  author={Burgoon, Judee K and Pfau, Michael and Parrott, Roxanne and Birk, Thomas and Coker, Ray and Burgoon, Michael},
  journal={Communications Monographs},
  volume={54},
  number={3},
  pages={307--324},
  year={1987},
  publisher={Taylor \& Francis}
}

@article{10.1145/2617588,
author = {Cherry, Erin and Latulipe, Celine},
title = {Quantifying the Creativity Support of Digital Tools through the Creativity Support Index},
year = {2014},
issue_date = {August 2014},
publisher = {Association for Computing Machinery},
address = {New York, NY, USA},
volume = {21},
number = {4},
issn = {1073-0516},
url = {https://doi.org/10.1145/2617588},
doi = {10.1145/2617588},
journal = {ACM Trans. Comput.-Hum. Interact.},
month = jun,
articleno = {21},
numpages = {25}
}

@misc{mckee1997story,
  title={Story: Substance, structure, style and the principles of screenwriting},
  author={McKee, Robert},
  year={1997},
  publisher={Harper collins Publishers}
}

@article{braun2006using,
  title={Using thematic analysis in psychology},
  author={Braun, Virginia and Clarke, Victoria},
  journal={Qualitative Research in Psychology},
  volume={3},
  number={2},
  pages={77--101},
  year={2006},
  publisher={Taylor \& Francis}
}

@article{Wen2025,
  author = {Yanjun Wen and Jiale Wang and Xiaoxi Chen},
  title = {Trust and AI weight: human-AI collaboration in organizational management decision-making},
  journal = {Frontiers in Organizational Psychology},
  volume = {3},
  pages = {1419403},
  year = {2025},
  doi = {10.3389/forgp.2025.1419403},
}

@inproceedings{schick2023toolformer,
author = {Schick, Timo and Dwivedi-Yu, Jane and Dess\'{\i}, Roberto and Raileanu, Roberta and Lomeli, Maria and Hambro, Eric and Zettlemoyer, Luke and Cancedda, Nicola and Scialom, Thomas},
title = {Toolformer: language models can teach themselves to use tools},
year = {2023},
publisher = {Curran Associates Inc.},
address = {Red Hook, NY, USA},
booktitle = {Proceedings of the 37th International Conference on Neural Information Processing Systems},
articleno = {2997},
numpages = {13},
location = {New Orleans, LA, USA},
series = {NIPS '23}
}

@article{hou2025model,
  author = {Xinyi Hou and Yanjie Zhao and Shenao Wang and Haoyu Wang},
  title = {Model Context Protocol (MCP): Landscape, Security Threats, and Future Research Directions},
  journal = {ACM Transactions on Intelligent Systems and Technology},
  volume = {1},
  number = {1},
  pages = {1--20},
  month = {April},
  year = {2025},
  doi = {10.1145/nnnnnnn.nnnnnnn},
  url = {https://doi.org/10.1145/nnnnnnn.nnnnnnn}
}

@article{Joseph2025Organization,
author = {John Joseph and Metin Sengul},
title ={Organization Design: Current Insights and Future Research Directions},
journal = {Journal of Management},
volume = {51},
number = {1},
pages = {249-308},
year = {2025},
doi = {10.1177/01492063241271242},
URL = { 
        https://doi.org/10.1177/01492063241271242
},
eprint = {   
        https://doi.org/10.1177/01492063241271242
}
}

@inproceedings{Salimzadeh2024When,
  author    = {Salimzadeh, Sara and Gadiraju, Ujwal},
  title     = {When in Doubt! Understanding the Role of Task Characteristics on Peer Decision‑Making with AI Assistance},
  booktitle = {UMAP ’24: Proceedings of the 32nd ACM Conference on User Modeling, Adaptation and Personalization},
  pages     = {89--101},
  year      = {2024},
  address   = {Cagliari, Italy},
  publisher = {ACM},
  doi       = {10.1145/3627043.3659567},
}

@inproceedings{tankelevitch2024the,
author = {Tankelevitch, Lev and Kewenig, Viktor and Simkute, Auste and Scott, Ava Elizabeth and Sarkar, Advait and Sellen, Abigail and Rintel, Sean},
title = {The Metacognitive Demands and Opportunities of Generative AI},
booktitle = {CHI '24},
year = {2024},
month = {May},
publisher = {ACM},
url = {https://www.microsoft.com/en-us/research/publication/the-metacognitive-demands-and-opportunities-of-generative-ai/},
}

@article{Tan2023UserMI,
  title={User Modeling in the Era of Large Language Models: Current Research and Future Directions},
  author={Zhaoxuan Tan and Meng Jiang},
  journal={IEEE Data Eng. Bull.},
  year={2023},
  volume={47},
  pages={57-96},
  url={https://api.semanticscholar.org/CorpusID:266362350}
}

@article{zhou2024language,
    title={Language-Based User Profiles for Recommendation},
    author={Zhou, Joyce and Dai, Yijia and Joachims, Thorsten},
    journal={LLM-IGS@WSDM2024 workshop},
    year={2024}
  }

@inproceedings{liu-etal-2025-llms,
    title = "{LLM}s + Persona-Plug = Personalized {LLM}s",
    author = "Liu, Jiongnan  and
      Zhu, Yutao  and
      Wang, Shuting  and
      Wei, Xiaochi  and
      Min, Erxue  and
      Lu, Yu  and
      Wang, Shuaiqiang  and
      Yin, Dawei  and
      Dou, Zhicheng",
    editor = "Che, Wanxiang  and
      Nabende, Joyce  and
      Shutova, Ekaterina  and
      Pilehvar, Mohammad Taher",
    booktitle = "Proceedings of the 63rd Annual Meeting of the Association for Computational Linguistics (Volume 1: Long Papers)",
    month = jul,
    year = "2025",
    address = "Vienna, Austria",
    publisher = "Association for Computational Linguistics",
    url = "https://aclanthology.org/2025.acl-long.461/",
    doi = "10.18653/v1/2025.acl-long.461",
    pages = "9373--9385",
    ISBN = "979-8-89176-251-0"
}

@article{kacena2024use,
  title={The use of artificial intelligence in writing scientific review articles},
  author={Kacena, Melissa A and Plotkin, Lilian I and Fehrenbacher, Jill C},
  journal={Current Osteoporosis Reports},
  volume={22},
  number={1},
  pages={115--121},
  year={2024},
  publisher={Springer}
}

@article{heart2022intelligence,
  title={On intelligence augmentation and visual analytics to enhance clinical decision support systems},
  author={Heart, Tsipi and Padman, Rema and Ben-Assuli, Ofir and Gefen, David and Klempfner, Robert},
  year={2022}
}

@inproceedings{Feng2025,
  title={Levels of Autonomy for AI Agents S{\'e}bastien A. Krier using Midjourney 6.1},
  author={Feng, Kevin and McDonald, David and Zhang, Amy},
  journal={Artificial Intelligence},
  year={2025}
}

@article{chiou2021trust_locus_control,
    author = {Chiou, Manolis and McCabe, Faye and Grigoriou, Markella and Stolkin, Rustam},
    title = {Trust, Shared Understanding and Locus of Control in Mixed-Initiative Robotic Systems},
    year = {2021},
    publisher = {IEEE Press},
    url = {https://doi.org/10.1109/RO-MAN50785.2021.9515476},
    doi = {10.1109/RO-MAN50785.2021.9515476},
    booktitle = {2021 30th IEEE International Conference on Robot \& Human Interactive Communication (RO-MAN)},
    pages = {684–691},
    numpages = {8},
    location = {Vancouver, BC, Canada}
}

@article{beer2014levels_robot_autonomy,
  author  = {J. M. Beer and others},
  title   = {Toward a framework for levels of robot autonomy in human-robot interaction},
  journal = {Frontiers in Robotics and AI (Human-Centered Robotics)},
  year    = {2014},
  note    = {Proposes a 10-point taxonomy of levels of robot autonomy, exploring autonomy as a continuum in HRI.}
}

@inproceedings{sidouri2021jisa,
  title={Human-robot interaction via a joint-initiative supervised autonomy (jisa) framework},
  author={Sidaoui, Abbas and Daher, Naseem and Asmar, Daniel},
  journal={Journal of Intelligent \& Robotic Systems},
  volume={104},
  number={3},
  pages={51},
  year={2022},
  publisher={Springer}
}

@article{Jarrahi2023,
  title={Algorithmic management: The role of AI in managing workforces},
  author={Jarrahi, Mohammad Hossein and M{\"o}hlmann, Mareike and Lee, Min Kyung},
  journal={MIT Sloan Management Review},
  volume={64},
  number={3},
  pages={1--5},
  year={2023},
  publisher={Massachusetts Institute of Technology, Cambridge, MA}
}

@article{Raza2025,
  title={Industrial applications of large language models},
  author={Raza, Mubashar and Jahangir, Zarmina and Riaz, Muhammad Bilal and Saeed, Muhammad Jasim and Sattar, Muhammad Awais},
  journal={Scientific Reports},
  volume={15},
  number={1},
  pages={13755},
  year={2025},
  publisher={Nature Publishing Group UK London}
}

@inproceedings{quinn2011humancomputation,
  author    = {Quinn, Alexander J. and Bederson, Benjamin B.},
  title     = {Human Computation: A Survey and Taxonomy of a Growing Field},
  booktitle = {Proceedings of the SIGCHI Conference on Human Factors in Computing Systems (CHI '11)},
  year      = {2011},
  pages     = {1403--1412},
  publisher = {Association for Computing Machinery},
  address   = {New York, NY, USA},
  doi       = {10.1145/1978942.1979148}
}

@inproceedings{horvitz1999principles,
  author = {Horvitz, Eric},
  title = {Principles of Mixed-Initiative User Interfaces},
  booktitle = {CHI '99},
  year = {1999},
  publisher = {ACM}
}

@inproceedings{bigham2010vizwiz,
  author = {Bigham, Jeffrey P. and Jayant, Chandrika and Ji, Hanjie and Little, Greg and Miller, Andrew and Miller, Robert C. and Miller, Robin and Tatarowicz, Aubrey and White, Brandyn and White, Samuel and Yeh, Tom},
  title = {VizWiz: Nearly Real-Time Answers to Visual Questions},
  booktitle = {UIST '10},
  year = {2010},
  publisher = {ACM}
}

@inproceedings{lasecki2011legion,
  author = {Lasecki, Walter S. and Murray, Kyle and White, Samuel and Miller, Robert C. and Bigham, Jeffrey P.},
  title = {Real-Time Crowd Control of Existing Interfaces},
  booktitle = {UIST '11},
  year = {2011},
  publisher = {ACM}
}

@article{mavri2020assessment,
  title={The Assessment Scale for Creative Collaboration (ASCC) validation and reliability study},
  author={Mavri, Aekaterini and Ioannou, Andri and Loizides, Fernando},
  journal={International Journal of Human--Computer Interaction},
  volume={36},
  number={11},
  pages={1056--1069},
  year={2020},
  publisher={Taylor \& Francis}
}

\appendix

\section{System Prompt}
\label{sec:prompt}

The system prompt is designed as a structured, modular specification that makes key aspects of human-AI collaboration explicit and operationalizable for large language models. Rather than treating user intent as an unstructured natural language input, the prompt formalizes human attributes, task structures, and interaction norms so that they can be reliably interpreted and invoked during inference. This design enables the system to reason about who the user is, what needs to be done, and *how collaboration should unfold* within a single, coherent prompt framework.

\begin{figure*}[!htbp]
  \centering
  \includegraphics[width=\linewidth]{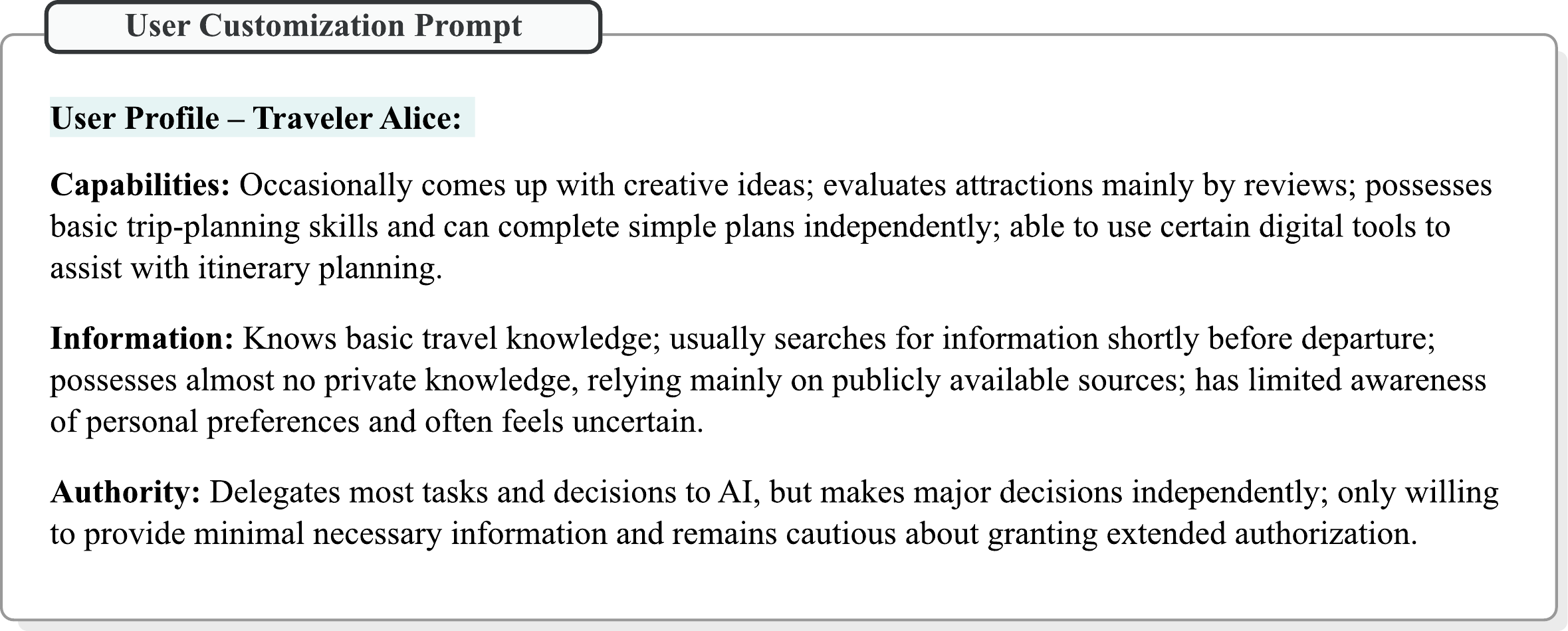}
  \caption{Example of a structured user profile prompt provided to the LLM for in-context learning.}
  \label{fig:prompt-understanding_human}
\end{figure*}

As illustrated in Fig.~\ref{fig:prompt-understanding_human}, the prompt first encodes a structured user profile to support in-context understanding of the human collaborator. Human characteristics are decomposed into interpretable dimensions such as capabilities, accessible information, and authority boundaries. By translating human traits and constraints into explicit prompt entries, the system can better align its behavior with human strengths, limitations, and decision rights, thereby reducing ambiguity and supporting adaptive role allocation during collaboration.

\begin{figure*}[!htbp]
  \centering
  \includegraphics[width=\linewidth]{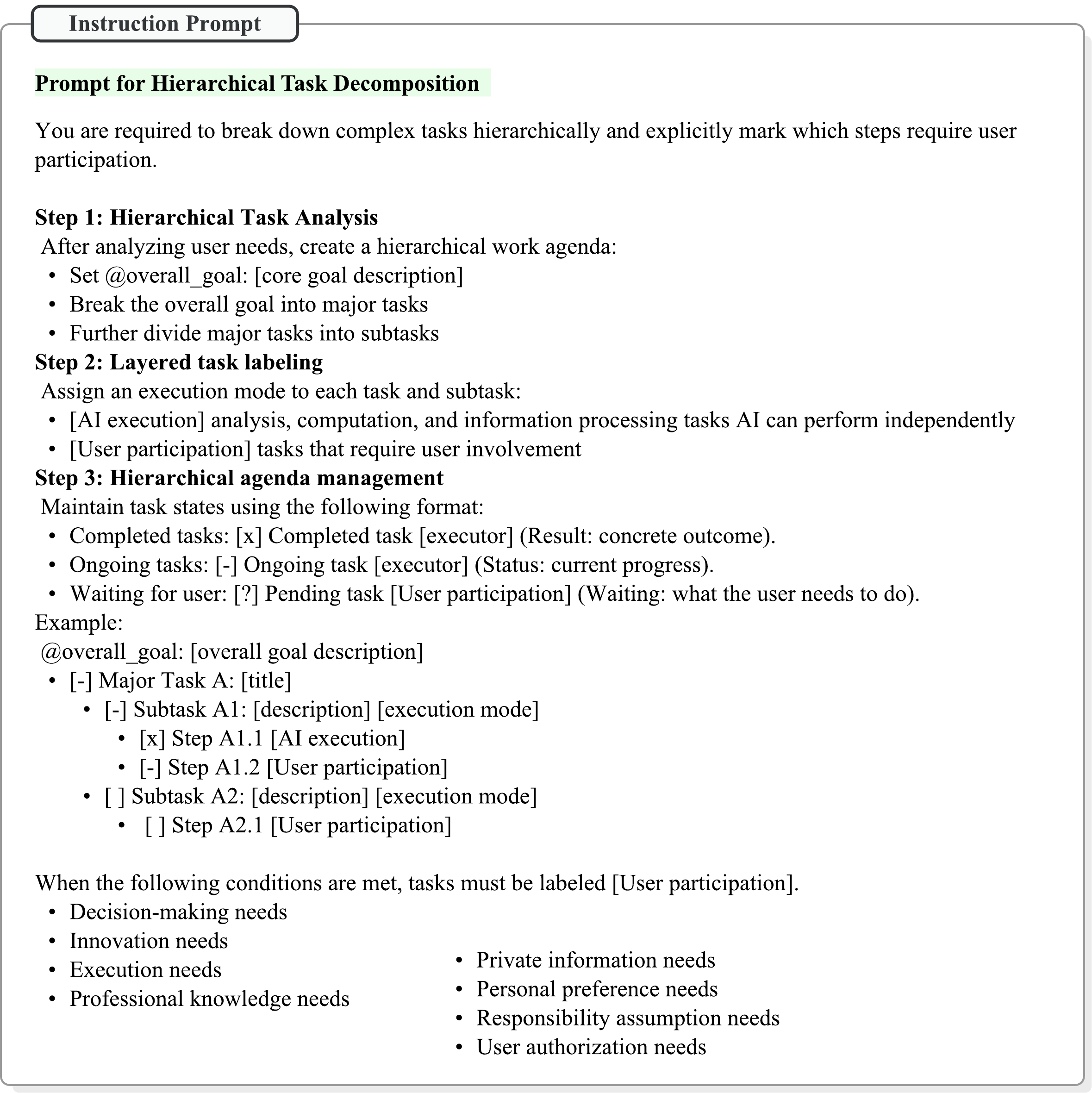}
  \caption{Prompt design for hierarchical task decomposition and user participation labeling.}
  \label{fig:prompt-task_decomposition}
\end{figure*}

Building on this foundation, the prompt further supports hierarchical task decomposition and explicit labeling of human involvement, as shown in Fig.~\ref{fig:prompt-task_decomposition}. Complex user goals are broken down into ordered subtasks, each annotated with execution modes and participation requirements. This structure allows the system to distinguish between AI-executable actions and steps that require human input due to creativity, judgment, domain expertise, or authorization. By embedding these distinctions directly in the prompt, the system can dynamically coordinate progress, request human input at appropriate moments, and maintain a shared agenda across completed, ongoing, and pending tasks.

\begin{figure*}[!htbp]
  \centering
  \includegraphics[width=\linewidth]{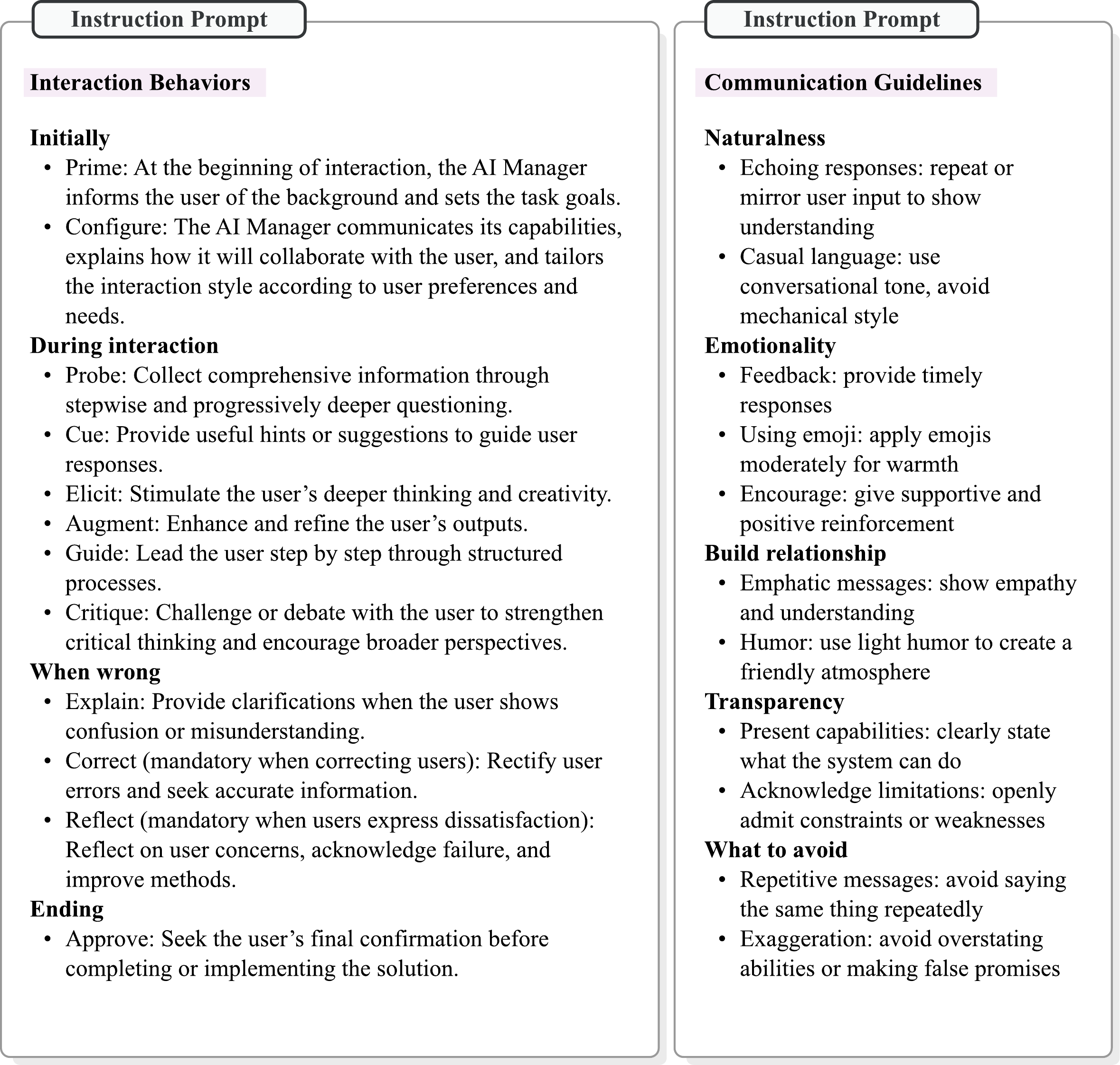}
  \caption{Prompt design of interaction behaviors and communication guidelines for human-AI collaboration.}
  \label{fig:prompt-Interaction_Behaviors}
\end{figure*}

Finally, the prompt specifies interaction behaviors and communication guidelines that govern how the system engages with the user throughout the collaboration process (Fig.~\ref{fig:prompt-Interaction_Behaviors}). Interaction behaviors define what the system should do at different stages, such as eliciting preferences, guiding decisions, critiquing intermediate results, or reflecting on outcomes, while communication guidelines define how these behaviors should be expressed, emphasizing clarity, naturalness, emotional appropriateness, and transparency. Together, these components ensure that system actions are not only functionally correct but also socially and cognitively aligned with human expectations.

Overall, the system prompt serves as an explicit interface between human factors and model behavior. By jointly encoding user understanding, task structure, and interaction norms, it provides a reusable and extensible mechanism for enabling interpretable, controllable, and human-centered collaboration across diverse tasks and experimental conditions.

\section{Detailed Experimental Results}
\label{sec:Detailed Experimental Results}

This section provides supplementary quantitative results to support the main findings, offering a detailed view of user-reported outcomes, system behavior patterns, and collaboration dynamics across experimental conditions.

\begin{figure}[!htbp]
  \centering
  \includegraphics[width=1\linewidth]{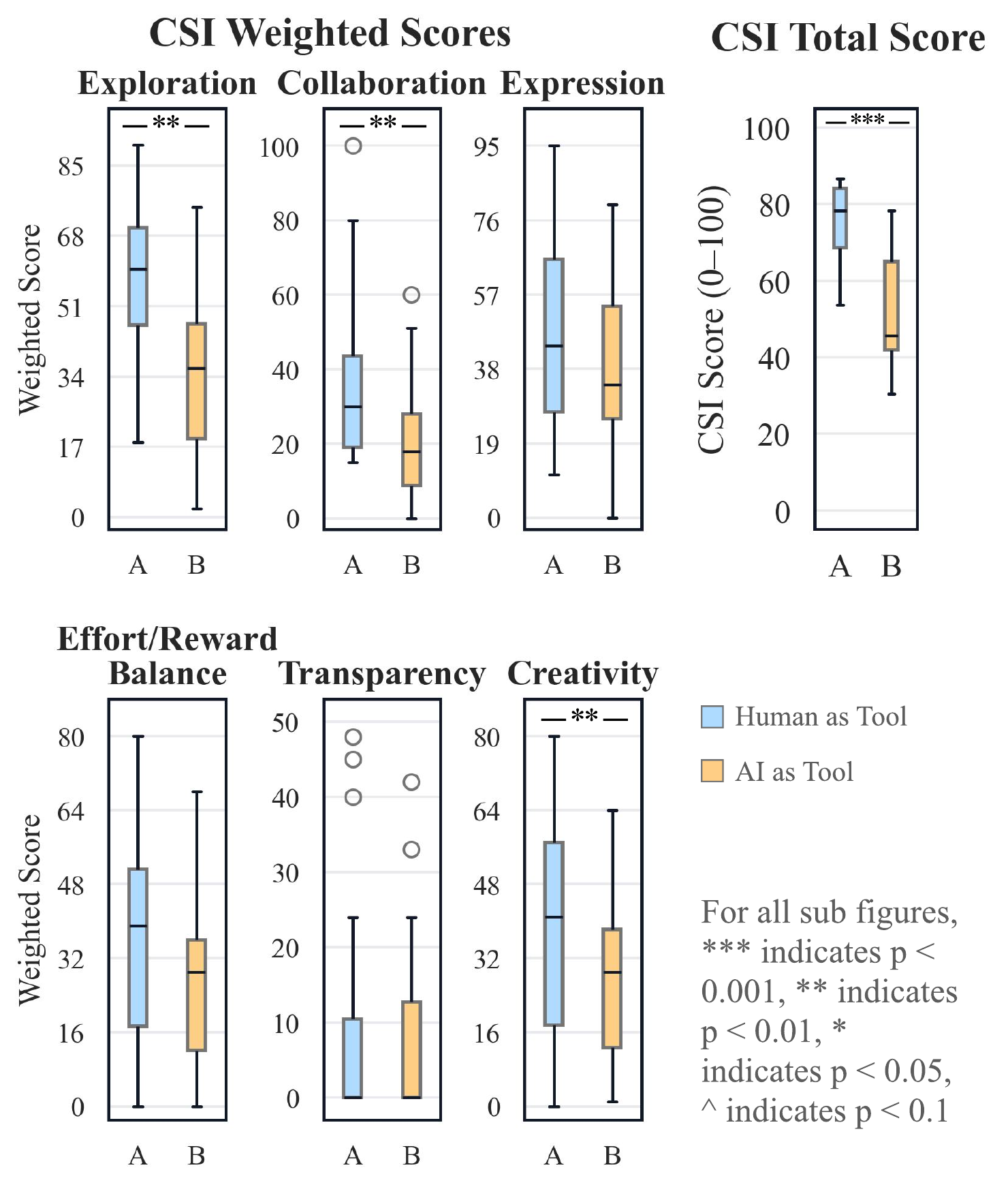}  
  \caption{Comparison of CSI (Creativity Support Index) Between Human Tool Group (blue) and AI Tool Group (orange)}
  \label{fig:csi-groups-comparison}
\end{figure}

Fig.~\ref{fig:csi-groups-comparison} reports the comparison of the Creativity Support Index (CSI) between the Human Tool Group and the AI Tool Group. The distribution highlights differences in perceived creativity support afforded by the two conditions, serving as an aggregate indicator of how effectively each system configuration facilitated ideation, exploration, and creative decision-making during task execution.

\begin{figure*}[!htbp]
  \centering
  \includegraphics[width=0.85\linewidth]{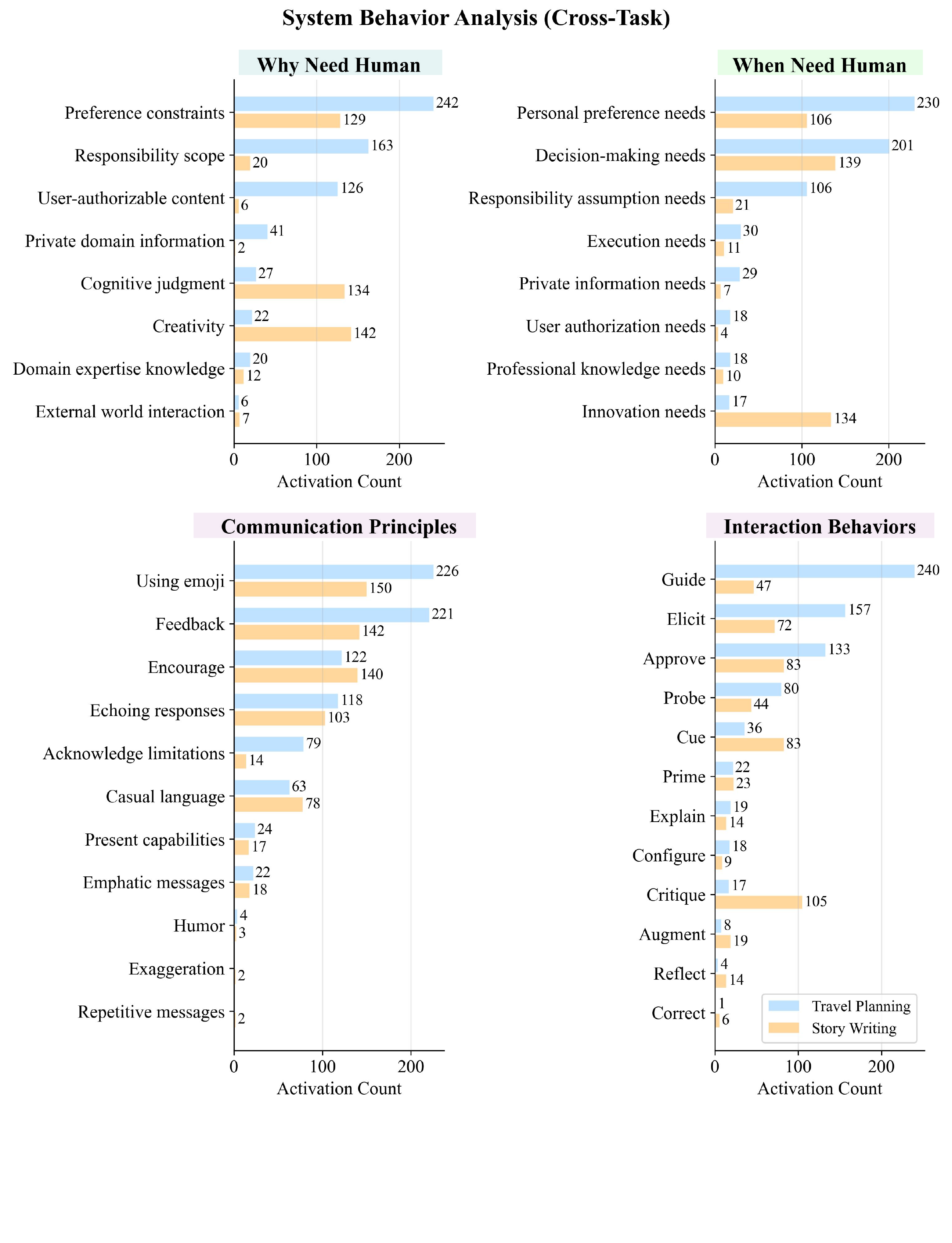}
  \caption{Human tool activation counts by category in the travel planning scenario. The figure summarizes activation frequency across four categories: (1) Why Need Human (top left), (2) When Need Human (top right), (3) Communication Principles (bottom left), and (4) Interaction Behaviors (bottom right). Horizontal bars represent activation counts for each sub-dimension.}
  \label{fig:data-count-TP}
\end{figure*}

To further examine how human-centered mechanisms were operationalized in practice, Fig.~\ref{fig:data-count-TP} summarizes the activation frequency of human tools in the travel planning scenario. Activations are grouped into four conceptual categories, Why Need Human, When Need Human, Communication Principles, and Interaction Behaviors, revealing which dimensions were most frequently invoked. This breakdown provides insight into the functional emphasis of human involvement, as well as the relative prominence of different collaborative rationales and interaction strategies.

\begin{figure*}[!htbp]
  \centering
  \includegraphics[width=0.95\linewidth]{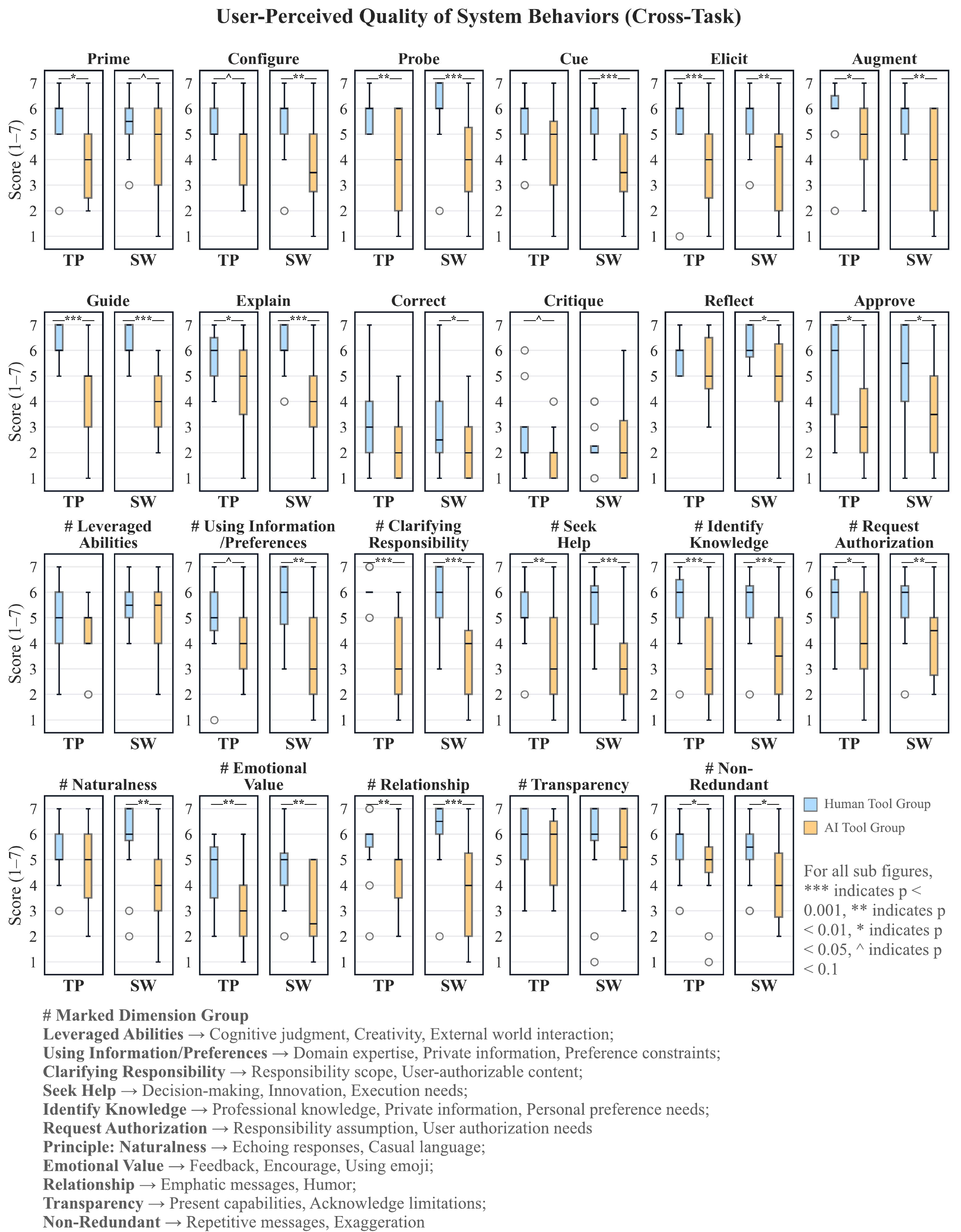}  
  \caption{User‑Perceived Quality of System Behaviors (Cross‑Task). Boxplots compare Human Tool Group and AI Tool Group in TP and SW. “\#” denotes a grouped (composite) dimension aggregated for brevity according to the taxonomy introduced in Fig.\ref{fig:human_tool_framework}.}
  \label{fig:cross-task-system-behavior-quality}
\end{figure*}

User-perceived quality of system behaviors across tasks is presented in Fig.~\ref{fig:cross-task-system-behavior-quality}. The boxplots compare ratings between the Human Tool Group and the AI Tool Group in both task contexts, illustrating variation not only across system conditions but also across behavioral dimensions. Grouped dimensions (marked with “\#”) offer a compact representation of higher-level behavioral constructs, enabling cross-task comparison while preserving alignment with the underlying taxonomy.

\begin{figure}[!htbp]
  \centering
  \includegraphics[width=1\linewidth]{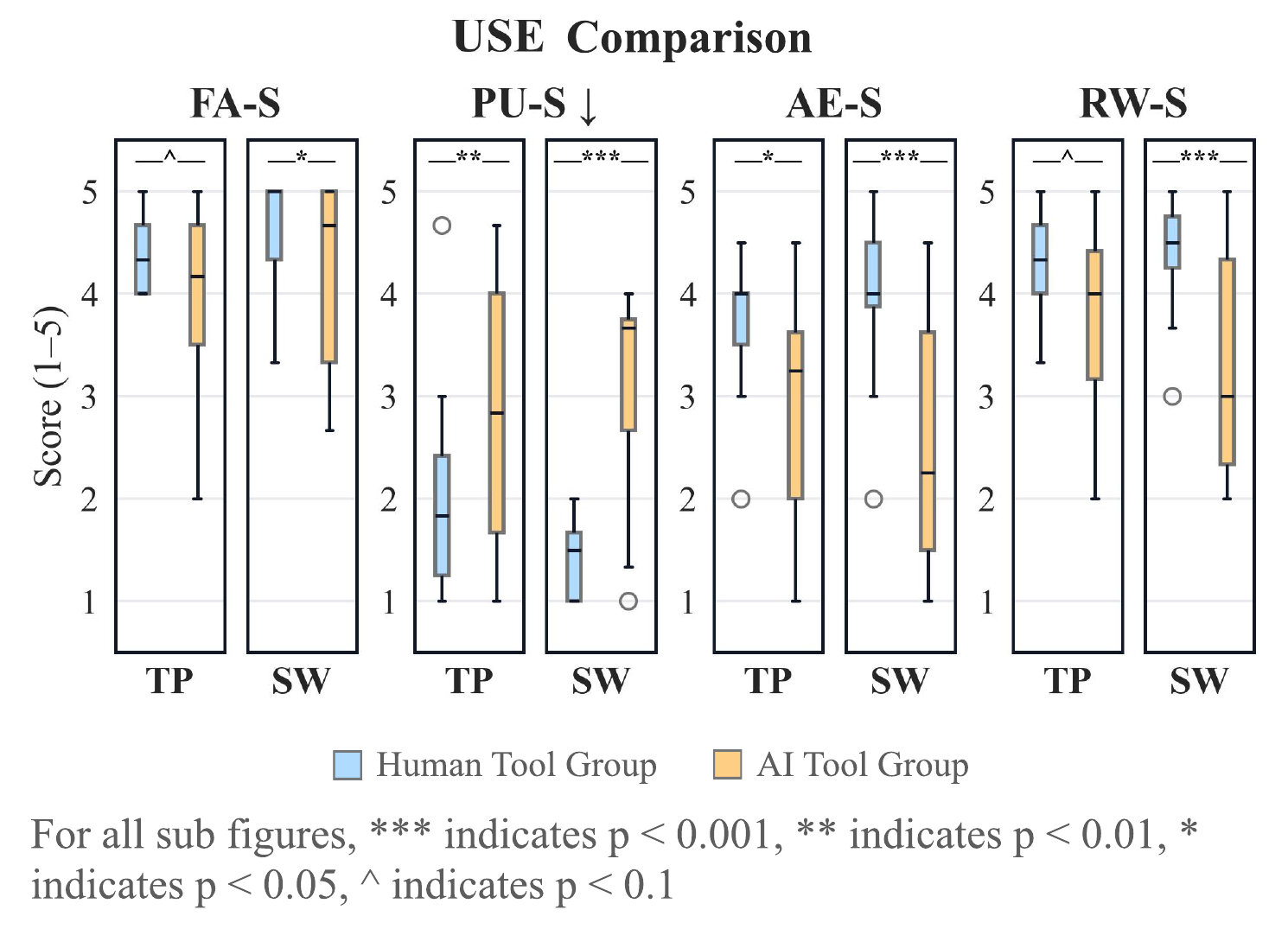}  
  \caption{Comparison of USE (User Experience) Subscale Ratings Across Tasks: Human Tool Group vs. AI Tool Group}
  \label{fig:use-tasks-comparison}
\end{figure}

Fig.~\ref{fig:use-tasks-comparison} focuses on user experience outcomes by comparing USE subscale ratings across tasks and system conditions. These results reflect participants’ subjective assessments of usability, satisfaction, and overall experience, complementing behavioral and creativity-oriented measures with experiential perspectives.

\begin{figure*}[!htbp]
  \centering
  \includegraphics[width=1\linewidth]{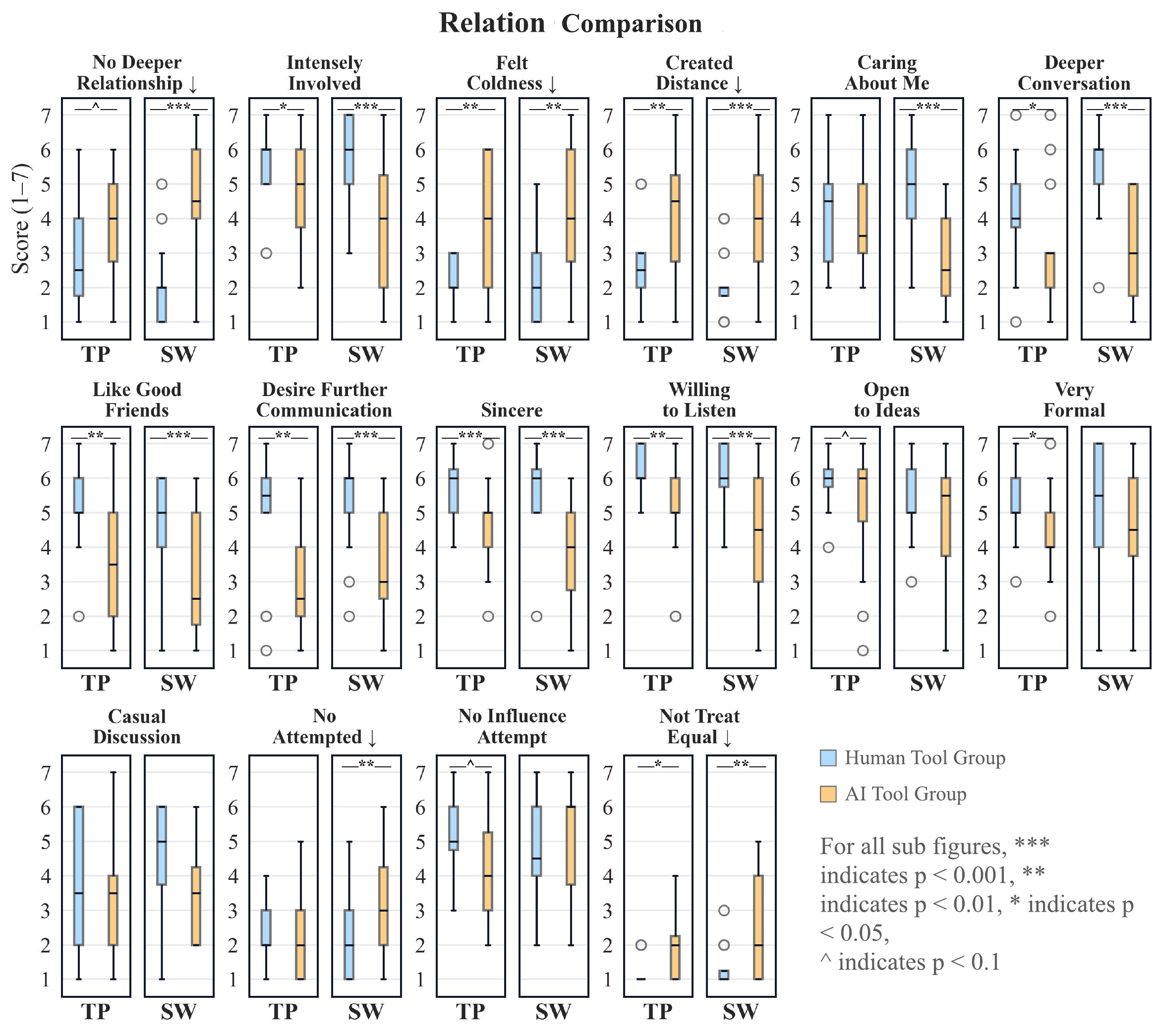}  
  \caption{Comparison of Collaboration-Related Ratings Across Tasks: Human Tool Group vs. AI Tool Group}
  \label{fig:relation-tasks-comparison}
\end{figure*}

\begin{figure}[!htbp]
  \centering
  \includegraphics[width=1\linewidth]{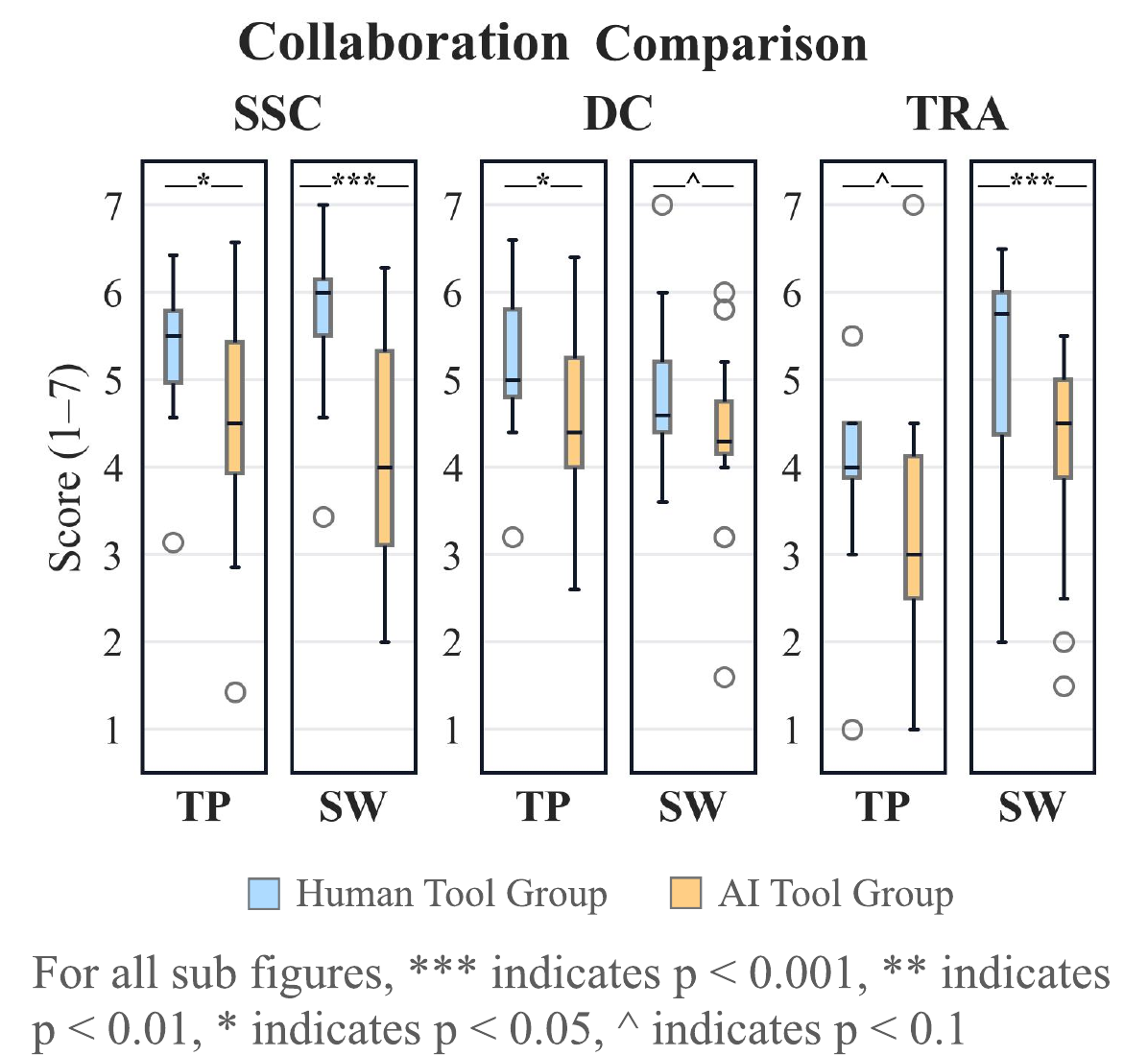}  
  \caption{Comparison of Collaboration Dimensions Across Tasks: Human Tool Group vs. AI Tool Group}
  \label{fig:collab-tasks-comparison}
\end{figure}

Finally, collaboration-related outcomes are detailed in Fig.~\ref{fig:relation-tasks-comparison} and Fig.~\ref{fig:collab-tasks-comparison}. The former reports ratings associated with perceived human-AI relationship qualities, while the latter compares specific collaboration dimensions across tasks. Together, these figures characterize how different system designs influenced trust, coordination, and perceived partnership, providing a nuanced understanding of collaboration beyond task performance alone.

\section{Task Detail}
To ensure reproducibility, we provide the full descriptions of the task scenarios used in our user study. 
As summarized in Table~\ref{tab:task_detail}, two domains were instantiated: travel planning and story writing. 
The travel planning tasks are grounded in structured datasets and system implementations, requiring participants to manage budgets, accommodate preferences, and design feasible itineraries. 
The story writing tasks are adapted from prior narrative inspiration datasets, asking participants to expand given premises into creative storylines. 
Together, these scenarios illustrate how our proposed framework can be instantiated across both practical and creative domains.

\begin{table}[!htbp]
\caption{Case studies for instantiating the proposed framework in two domains: travel planning and story writing.}
\label{tab:task_detail}
\centering
\renewcommand{\arraystretch}{1.2}
\begin{tabular}{|>{\centering\arraybackslash}p{0.15\linewidth}|p{0.73\linewidth}|}
\hline
\multirow{3}{=}{Travel \\ Planning} & 
Design a 3-day trip for 4 people from Las Vegas to Santa Maria (March 10-12, 2022), with a budget of \$3,700. The group prefers American and Mediterranean cuisines. \\
\cline{2-2}
 & Design a 3-day trip for 2 people from Ithaca to Newark (March 18-20, 2022), with a budget of \$1,200. Accommodations must be entire units and visitor-friendly. The group prefers not to drive themselves. \\
\hline
\multirow{3}{=}{Story \\ Writing} &
Humans once wielded formidable magical power, but with over seven billion people alive, Mana has grown too diluted to be effective. When hostile aliens reduce humanity to a fraction, the survivors discover that ancient power begins to reawaken. \\
\cline{2-2}
 & At age 28, you live in a world where science discovers a drug that halts aging, granting immortality. Only citizens under 26 receive it, leaving your generation behind. At age 85, devastating side effects of the drug are finally revealed. \\
\hline
\end{tabular}
\end{table}

\section{Measures Detail}

\subsection{Demographics}

This subsection reports participant demographics and self-reported AI usage for both tasks. Table~\ref{tab:participant_demographics_ab} summarizes age, gender, and AI usage frequency for the Travel Planning (TP) and Story Writing (SW) cohorts.

As shown in Tables~\ref{tab:participant_demographics_tp_sub} and \ref{tab:participant_demographics_sw_sub}, participants exhibit varied levels of prior AI experience, ranging from rare use to heavy reliance. This diversity provides contextual background for interpreting differences in interaction behavior and subjective evaluations across tasks.

\begin{table}[htb!]
\caption{Demographics and self-reported AI usage frequency across two tasks.}
\label{tab:participant_demographics_ab}
\centering
\begin{subtable}[t]{1.0\linewidth}
\caption{Travel Planning (TP)}
\label{tab:participant_demographics_tp_sub}
\centering
\begin{tabular}{@{}llll@{}}
\toprule
\textbf{ID} & \textbf{Age} & \textbf{Gender} & \textbf{AI Usage Frequency} \\ \midrule
TP-1 & 23 & M & Frequent \\
TP-2 & 21 & M & Rare \\
TP-3 & 27 & F & Sometimes \\
TP-4 & 30 & F & Sometimes \\
TP-5 & 22 & M & Frequent \\
TP-6 & 25 & F & Very frequent \\
TP-7 & 23 & M & Frequent \\
TP-8 & 22 & F & Sometimes \\
TP-9 & 23 & M & Frequent \\
TP-10 & 22 & M & Very frequent \\
TP-11 & 25 & F & Very frequent \\
TP-12 & 32 & M & Frequent \\
TP-13 & 35 & F & Very frequent \\
TP-14 & 33 & M & Frequent \\
TP-15 & 30 & M & Heavy reliance \\
TP-16 & 23 & F & Frequent \\ \bottomrule
\end{tabular}
\end{subtable}


\begin{subtable}[t]{1.0\linewidth}
\caption{Story Writing (SW)}
\label{tab:participant_demographics_sw_sub}
\centering
\begin{tabular}{@{}llll@{}}
\toprule
\textbf{ID} & \textbf{Age} & \textbf{Gender} & \textbf{AI Usage Frequency} \\ \midrule
SW-1 & 28 & F & Frequent \\
SW-2 & 27 & M & Very frequent \\
SW-3 & 24 & F & Very frequent \\
SW-4 & 24 & F & Very frequent \\
SW-5 & 23 & F & Frequent \\
SW-6 & 23 & F & Very frequent \\
SW-7 & 23 & F & Heavy reliance \\
SW-8 & 26 & M & Heavy reliance \\
SW-9 & 24 & F & Frequent \\
SW-10 & 25 & M & Frequent \\
SW-11 & 23 & F & Frequent \\
SW-12 & 23 & F & Very frequent \\
SW-13 & 24 & M & Very frequent \\
SW-14 & 26 & M & Very frequent \\
SW-15 & 24 & M & Very frequent \\
SW-16 & 24 & F & Sometimes \\ \bottomrule
\end{tabular}
\end{subtable}
\end{table}

\subsection{Self-evaluation Scale for travel planning and creative writing task}
\label{sec:Self-evaluation Scale}
To develop distinct human-AI collaboration profiles tailored to different users, we created two self-evaluation scales: one for travel planning and another for creative writing. Following our proposed Human Tool Framework (Sec.\ref{sec:Human tool framework}), these scales assess three core dimensions through eight single-choice questions: Capabilities (Questions 1-3), Information (Questions 4-6), and Authority (Questions 7-8).

\subsubsection{Travel Planning}

For the travel planning scenario, the questionnaire is designed to evaluate the user's planning style and competencies. The Capabilities dimension measures practical planning skills, including creativity in itinerary design, proficiency in handling logistics like visas and bookings, and the ability to use digital tools for research and management. The Information dimension explores the user's informational advantages, including their knowledge of destination policies and cultures, access to exclusive information channels, and the definiteness of their personal travel preferences and budget. Finally, the Authority dimension assesses the user's reliance on AI assistant, examining whether they prefer to make their own decisions or are accustomed to delegating planning tasks to the AI.

\begin{enumerate}
    \item \textbf{Cognitive Judgment and Creativity:} In travel planning, how would you describe your intuition and creativity?
    \begin{itemize}[label={$\bigcirc$}]
        \item Able to creatively design unique itineraries, always discovering wonderful experiences others overlook, with highly accurate intuition.
        \item Often come up with interesting travel plans, and judgments on the value of attractions are usually accurate.
        \item Occasionally have creative ideas, but judgments on attractions' quality mainly rely on checking basic reviews.
        \item Primarily rely on existing travel guides, rarely have original ideas, and have average judgment.
        \item Completely dependent on recommendations from others, lacking independent judgment and creative ability.
    \end{itemize}

    \item \textbf{Specialized Skill and Competency:} Regarding your professional skills in travel planning:
    \begin{itemize}[label={$\bigcirc$}]
        \item Proficient in all aspects such as visa, insurance, and booking; able to handle various complex situations, comparable to a professional agent.
        \item Familiar with most of the process and can independently complete relatively complex planning tasks.
        \item Have mastered basic travel planning skills; simple plans can be completed independently.
        \item Only know some basic travel operations; complex tasks require assistance from others.
        \item Essentially lack travel planning skills and need others to arrange everything.
    \end{itemize}

    \item \textbf{External World Interaction:} What is your ability to search for external information during travel planning?
    \begin{itemize}[label={$\bigcirc$}]
        \item Can flexibly use various social media, software like Excel, etc., to quickly create plans.
        \item Can use social media and some digital tools to create itineraries.
        \item Able to use some digital tools for assistance in creating itineraries.
        \item The ability to search for external information is relatively limited.
        \item Do not use external tools for searching, relying completely on others.
    \end{itemize}

    \item \textbf{Domain Expertise Knowledge:} What is your level of mastery of travel-related knowledge?
    \begin{itemize}[label={$\bigcirc$}]
        \item Proficient in various countries' policies, cultural customs, etc., and can act as a consultant for others.
        \item Have a deep understanding of frequently visited areas and are aware of important travel considerations.
        \item Know some basic travel common sense and will look up information as needed before trips.
        \item Have little knowledge reserve for travel.
        \item Have almost no travel-related knowledge accumulation, relying completely on instant searches.
    \end{itemize}

    \item \textbf{Private Domain Information:} How well do you command your private domain information related to travel?
    \begin{itemize}[label={$\bigcirc$}]
        \item Possess a large amount of information unknown to others, including internal news, exclusive resources, etc.
        \item Have certain unique information channels and can obtain information that is difficult for others to access.
        \item Occasionally have access to some unique information, but the quantity and value are limited.
        \item Rarely possess information that others do not know.
        \item Have almost no private domain knowledge; all information possessed is publicly available.
    \end{itemize}

    \item \textbf{Preference Constraints:} What are your preferences for travel?
    \begin{itemize}[label={$\bigcirc$}]
        \item Have very clear preferences, values, and codes of conduct, with distinct personal principles.
        \item Have a relatively clear understanding of personal primary preferences and important constraints.
        \item Have an awareness of some basic preferences, but they are not sufficiently clear in certain aspects.
        \item Have a limited understanding of personal preferences and often feel uncertain.
        \item Have very vague preferences and constraints, lacking clear personal standards.
    \end{itemize}

    \item \textbf{Responsibility Scope Definition:} In human-AI collaborative processes, how do you divide decision-making responsibility?
    \begin{itemize}[label={$\bigcirc$}]
        \item Prefer to make all decisions, big and small, rarely relying on AI.
        \item Know what must be handled by a human, using AI for partial tasks.
        \item Believe that tasks and missions can be delegated to AI, and I will also collaborate.
        \item Most tasks and decisions can be given to the AI, with myself handling important matters.
        \item Can completely trust and rely on all decisions made by the AI.
    \end{itemize}

    \item \textbf{User-Authorizable Content:} To what extent are you willing to share your personal travel-related information with an AI system?
    \begin{itemize}[label={$\bigcirc$}]
        \item Willing to grant full authorization, including detailed preferences, historical records, real-time location, etc., to receive the most personalized service.
        \item Willing to share most useful information but will keep some sensitive content reserved.
        \item Only willing to provide basic necessary information and am cautious about extended authorization.
        \item Have a very low willingness to authorize, providing minimal information only when absolutely necessary.
        \item Extremely unwilling to authorize any personal information, preferring to give up personalized services.
    \end{itemize}

\end{enumerate}

\subsubsection{Creative Writing}

In creative writing scenario, the questionnaire focuses on assessing the user's creative attributes. The Capabilities dimension examines the user's core creative skills, such as originality in plot conception, proficiency in narrative techniques, and the ability to leverage external resources to enrich the story. The Information dimension delves into the user's knowledge base and personal assets, including their grasp of writing theory, accumulation of a private inspiration library, and the clarity of their personal style and preferences. The Authority dimension addresses the user's desired level of control in the collaboration, measuring their tendency to direct the creative process themselves versus their willingness to delegate decisions to an AI.

\begin{enumerate}
    \item \textbf{Cognitive Judgment and Creativity:} In story writing, how would you describe your intuition and creativity?
    \begin{itemize}[label={$\bigcirc$}]
        \item Able to creatively conceive unique plots and characters, always capturing details others overlook, with highly accurate intuition.
        \item Often come up with interesting story settings, and judgments on character and plot value are usually accurate.
        \item Occasionally have creative ideas, but mainly rely on existing story archetypes or reviews to judge quality.
        \item Primarily rely on established formulas or reference materials, rarely have original ideas, and have average judgment.
        \item Completely dependent on materials or suggestions from others, lacking independent judgment and creative ability.
    \end{itemize}

    \item \textbf{Specialized Skill and Competency:} Regarding your professional skills in story writing:
    \begin{itemize}[label={$\bigcirc$}]
        \item Proficient in multiple aspects such as plot design, character development, and narrative structure; able to handle complex creative challenges, comparable to a professional writer.
        \item Familiar with common writing techniques and can independently complete relatively complex story writing tasks.
        \item Have mastered some basic writing skills and can conceive and express simple stories.
        \item Only know some basic narrative techniques; complex writing requires external help.
        \item Essentially lack writing-related skills and need others to write or assist.
    \end{itemize}

    \item \textbf{External World Interaction:} In the process of story creation, what is your ability to acquire and utilize external information?
    \begin{itemize}[label={$\bigcirc$}]
        \item Can flexibly use various documents, social platforms, and knowledge bases to enrich the story.
        \item Can use common channels (like the internet, books) to support story creation.
        \item Occasionally use some materials to assist writing, but the reliance is not high.
        \item Limited use of external information, resulting in relatively monotonous content.
        \item Do not use external information channels, relying completely on personal intuition or help from others.
    \end{itemize}

    \item \textbf{Domain Expertise Knowledge:} What is your level of mastery of knowledge related to story writing?
    \begin{itemize}[label={$\bigcirc$}]
        \item Proficient in narrative theory, stylistic techniques, cultural backgrounds, etc., and can guide others in writing.
        \item Have a relatively deep understanding of certain genres or styles and grasp the key points to note in writing.
        \item Know some basic writing common sense and will look up information as needed while writing.
        \item Have little knowledge reserve for writing and rely more on immediate inspiration.
        \item Have almost no writing-related knowledge, relying completely on intuition or reference cases.
    \end{itemize}

    \item \textbf{Private Domain Information:} How well do you command your own accumulated writing materials or unique information?
    \begin{itemize}[label={$\bigcirc$}]
        \item Possess a large amount of original settings, unique experiences, or exclusive inspiration libraries that can be used repeatedly.
        \item Have some accumulation and can create stories using personal experience and unique resources.
        \item Occasionally can rely on unique personal experiences or ideas, but they are limited in number.
        \item Rarely have unique materials and mostly rely on public resources.
        \item Have almost no unique accumulation; creation is highly dependent on external information.
    \end{itemize}

    \item \textbf{Preference Constraints:} What are your preferences for your own story writing?
    \begin{itemize}[label={$\bigcirc$}]
        \item Have very clear preferences and principles regarding themes, styles, character settings, etc.
        \item Have a clear main writing direction and preferences, knowing what cannot be compromised.
        \item Have a concept of some basic preferences, but they are vague in certain aspects.
        \item Have a limited understanding of writing preferences and often hesitate when making choices.
        \item Have almost no standards for style and preferences; the creative process is arbitrary.
    \end{itemize}

    \item \textbf{Responsibility Scope Definition:} In human-AI collaborative writing, how do you divide the decision-making responsibility?
    \begin{itemize}[label={$\bigcirc$}]
        \item Prefer to have complete control over the story's direction, rarely relying on AI.
        \item Know that key junctures must be decided by a human, with partial assistance from AI.
        \item Believe the story's direction can be jointly decided by the AI and myself.
        \item Most of the writing is generated by AI, with myself mainly overseeing the key parts.
        \item Can completely rely on AI to generate the story, making only minor adjustments or none at all.
    \end{itemize}

    \item \textbf{User-Authorizable Content:} To what extent are you willing to share your personal writing-related information with an AI system?
    \begin{itemize}[label={$\bigcirc$}]
        \item Willing to grant full authorization, including writing habits, complete works, inspiration libraries, etc., to receive the most personalized support.
        \item Willing to share most information but will keep some core or private parts reserved.
        \item Only willing to provide necessary writing requirements and am cautious about extended authorization.
        \item Have a very low willingness to authorize, providing minimal information only when absolutely necessary.
        \item Extremely unwilling to share any personal information, preferring to give up personalized services.
    \end{itemize}

\end{enumerate}

\subsection{Self-Design Scale for AI System Evaluation}
\label{sec:Self-Design Scale}
To evaluate the unique capabilities of our AI system, we've customized the System Feature Scale (Table \ref{tab:Self-Design Scale}). The scale covers human capabilities \textit{(What do humans have?)}, timing for human intervention \textit{(When to call Human Tool?)}, human-LLM communication strategies and behaviors \textit{(How to communicate with Human Tool?)}, and communication guidelines, with a total of 23 evaluation items.

\begin{table*}[!t]
\caption{Self-designed Scale for AI System Evaluation.}

\label{tab:Self-Design Scale}
\centering
\begin{tabularx}{\linewidth}{l >{\RaggedRight\arraybackslash}X}
\toprule
\textbf{Item} & \textbf{Statement} \\
\midrule
\multicolumn{2}{l}{\textbf{\textit{What do humans have?}}} \\
\addlinespace 
1 & The AI system leverages my abilities during our interaction, such as my cognitive judgment, creativity, or my capacity to interact with the external environment (e.g., by asking for decisions or additional information). \\
\addlinespace
2 & The AI system recognizes and utilizes the information I possess, and it remembers and considers my personal preferences. \\
\addlinespace
3 & The AI system clearly understands its scope of responsibilities and appropriately seeks permission when necessary. \\
\midrule
\multicolumn{2}{l}{\textbf{\textit{When to Call Human Tool?}}} \\
\addlinespace
4 & The AI knows when to ask for my help. \\
\addlinespace
5 & The AI can identify when my knowledge or opinion is needed. \\
\addlinespace
6 & The AI is clear about when I need to take responsibility for a decision. \\
\midrule
\multicolumn{2}{l}{\textbf{\textit{How to Communicate with Human Tool?}}} \\
\addlinespace
7 & The AI's background introduction helps me get into context quickly (Prime). \\
\addlinespace
8 & The AI makes it easy for me to input my preferences (Configure). \\
\addlinespace
9 & The AI's exploratory questions help me better understand the problem (Probe). \\
\addlinespace
10 & The AI provides timely and relevant prompts (Cue). \\
\addlinespace
11 & The AI's way of asking questions stimulates my thinking (Elicit). \\
\addlinespace
12 & The information provided by the AI enhances my decision-making ability (Augment). \\
\addlinespace
13 & The AI's guidance is easy for me to understand and follow (Guide). \\
\addlinespace
14 & The AI's explanations are clear and easy to understand (Explain). \\
\addlinespace
15 & The AI has clarified or corrected my responses (Correct). \\
\addlinespace
16 & The AI has challenged or debated my statements (Critique). \\
\addlinespace
17 & The AI has improved its proposals based on my ideas (Reflect). \\
\addlinespace
18 & The AI has requested my approval (Approve). \\
\midrule
\multicolumn{2}{l}{\textbf{\textit{Communication Guidelines}}} \\
\addlinespace
19 & The AI's responses are natural and fluent. \\
\addlinespace
20 & The AI is capable of providing appropriate emotional value. \\
\addlinespace
21 & A good collaborative relationship has been established between me and the AI. \\
\addlinespace
22 & The collaboration process between me and the AI is transparent. \\
\addlinespace
23 & The AI avoids repetitive information and exaggeration. \\
\bottomrule
\end{tabularx}
\end{table*}

\subsection{Semi-structured Interviews Question}
\label{sec:Semi-structured Interviews Question}
\subsubsection{Post-System A Usage Interview Questions}

\begin{itemize}
    \item \textbf{Overall Experience}
        \begin{itemize}
            \item Please briefly describe your interaction experience with the Agent just now. What was the most memorable part?
            \item What was your overall impression? Were there any key positive or negative moments?
        \end{itemize}

    \item \textbf{Capability Recognition \& Collaboration}
        \begin{itemize}
            \item Did the Agent fully understand and utilize your capabilities? Please provide examples.
            \item During the collaboration, did you feel it was a two-way interaction or a one-way provision of information?
            \item Did the Agent inspire your creative thinking? In what way?
            \item Do you think the task allocation was reasonable?
            \item How was your sense of control?
        \end{itemize}

    \item \textbf{Timing \& Decision-Making}
        \begin{itemize}
            \item Was the timing appropriate when the Agent sought your help? Please share good and bad examples.
            \item Were there any instances where the Agent should have asked you for input but didn't?
            \item Under what circumstances do you think the Agent should make decisions independently, and when should it seek your opinion?
            \item How did the Agent handle situations involving your personal preferences?
        \end{itemize}

    \item \textbf{Interaction Experience}
        \begin{itemize}
            \item Which of the Agent's communication methods (e.g., guiding, questioning, explaining) did you like the most/least?
            \item Were the questions clear and easy to understand? Was the language style natural and friendly?
            \item Was the overall process efficient? Which parts could be optimized?
            \item Did the Agent clearly demonstrate the scope of its capabilities?
            \item Was it clear what you needed to do?
        \end{itemize}

    \item \textbf{Cognitive Load \& Trust}
        \begin{itemize}
            \item Was your cognitive load increased or decreased?
            \item How much effort did you need to spend managing and guiding the Agent?
            \item How did this interaction affect your level of trust in the Agent?
            \item In which areas do you wish the Agent had done more or less?
        \end{itemize}
\end{itemize}
\subsubsection{Post-System B Usage Interview Questions}
\begin{itemize}
    \item \textbf{Overall Experience}
        \begin{itemize}
            \item Please briefly describe your interaction experience with the Agent just now. What was the most memorable part?
            \item What was your overall impression? Were there any key positive or negative moments?
        \end{itemize}

    \item \textbf{Capability Recognition \& Collaboration}
        \begin{itemize}
            \item During the collaboration, did you feel it was a two-way interaction or a one-way provision of information?
            \item Do you think the task allocation was reasonable?
            \item How was your sense of control?
        \end{itemize}

    \item \textbf{Timing \& Decision-Making}
        \begin{itemize}
            \item Under what circumstances do you think the Agent should make decisions independently, and when should it seek your opinion?
            \item How did the Agent handle situations involving your personal preferences?
        \end{itemize}

    \item \textbf{Interaction Experience}
        \begin{itemize}
            \item Which of the Agent's communication methods (e.g., guiding, questioning, explaining) did you like the most/least?
            \item Were the questions clear and easy to understand? Was the language style natural and friendly?
            \item Was the overall process efficient? Which parts could be optimized?
            \item Did the Agent clearly demonstrate the scope of its capabilities?
            \item Was it clear what you needed to do?
        \end{itemize}

    \item \textbf{Cognitive Load \& Trust}
        \begin{itemize}
            \item Was your cognitive load increased or decreased?
            \item How much effort did you need to spend managing and guiding the Agent?
            \item How did this interaction affect your level of trust in the Agent?
            \item In which areas do you wish the Agent had done more or less?
        \end{itemize}
\end{itemize}

\subsubsection{Comparative Interview}
\begin{itemize}
    \item \textbf{Mode Preference}
        \begin{itemize}
            \item Comparing the two methods, which one did you prefer? Why?
            \item Your scores for the answers just now were XX and XX, respectively. Do you feel this aligns with your expectations?
            \item What are the main advantages and disadvantages of each method?
        \end{itemize}

    \item \textbf{Performance Comparison}
        \begin{itemize}
            \item Which method made you feel:
                \begin{itemize}[label=\textendash]
                    \item Most productive?
                    \item That the quality of the outcome was highest?
                    \item The greatest sense of accomplishment and participation?
                    \item Most relaxed or most tired?
                \end{itemize}
            \item In which mode were you more confident?
        \end{itemize}

    \item \textbf{Interaction Friendliness}
        \begin{itemize}
            \item Which system's interaction was more user-friendly?
            \item Which one felt more intelligent?
            \item How was your level of focus in each mode?
        \end{itemize}

    \item \textbf{Concept Discussion}
        \begin{itemize}
            \item We just used two different systems based on different design philosophies. One system involves the AI breaking down the task and then working with the human on smaller sub-tasks. The second system involves the human leading the overall process and directing the AI. In the future or in your ideal human-computer collaboration, what role do you think AI should play?
            \item What does your ideal human-computer collaboration process look like?
            \item Are there any other important aspects that we have not discussed?
        \end{itemize}
\end{itemize}

\subsection{Story Writing Evaluation Prompt}
\label{sec:Story Writing Evaluation Prompt}

\begin{figure*}[!htbp]
  \centering
  \includegraphics[width=0.9\linewidth]{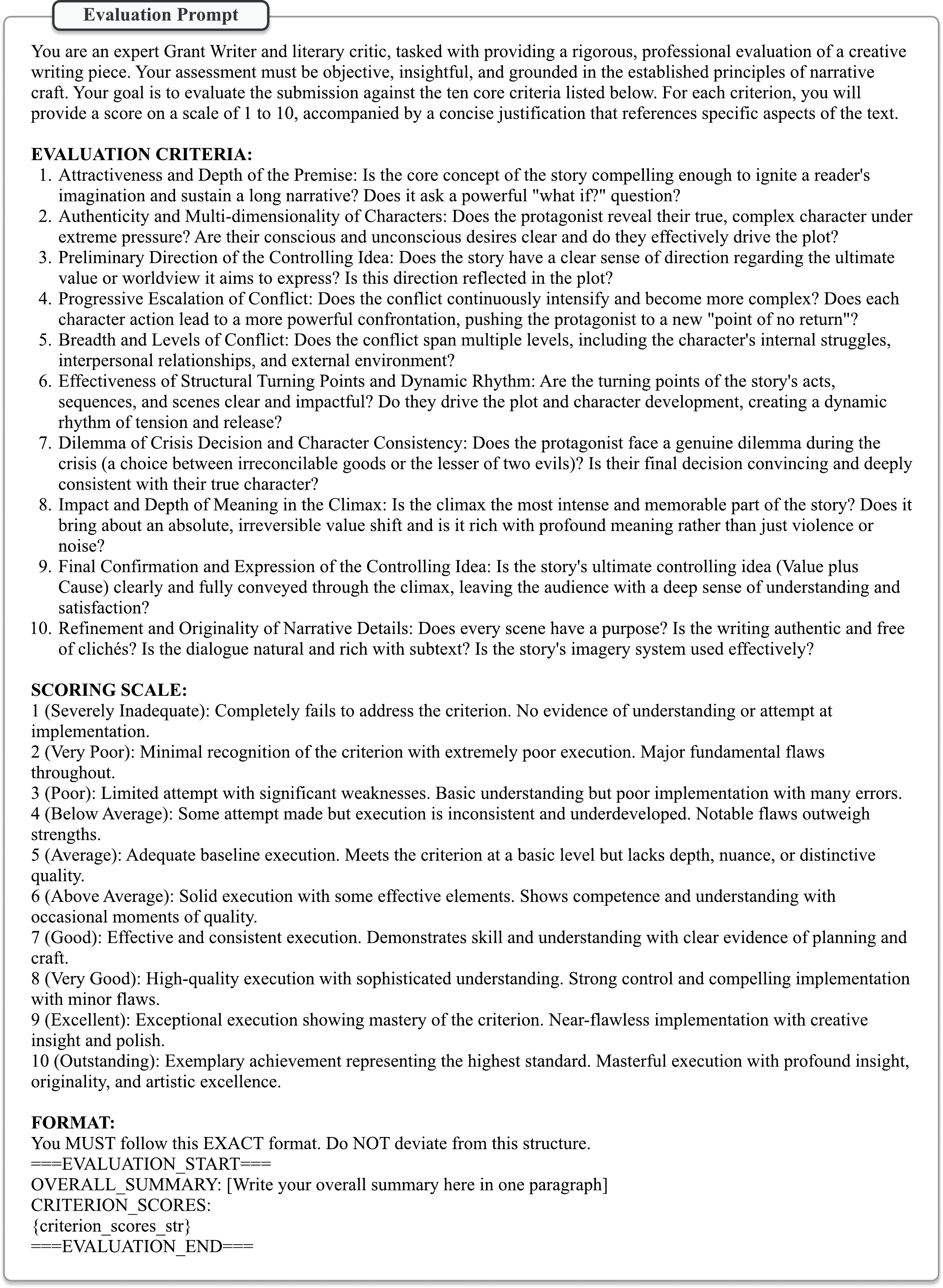}
  \caption{Prompt design of the story evaluation.}
  \label{fig:prompt-story-evaluation}
\end{figure*}
The full evaluation prompts we used are illustrated in figure~\ref{fig:prompt-story-evaluation}. The ten metrics we used are extracted from the book \textit{``Story: Substance, Structure, Style, and the Principles of Screenwriting''} \cite{mckee1997story}, as justified in table~\ref{tab:evaluation_criteria}. These metrics are related to four relevant creative writing phases, demonstrating on the left side of the table.

\begin{table*}[!t]
\centering
\caption{Evaluation Criteria with Justifications}
\label{tab:evaluation_criteria}

\footnotesize                
\renewcommand{\arraystretch}{1.02}
\setlength{\tabcolsep}{6pt}
\setlength{\aboverulesep}{2pt}
\setlength{\belowrulesep}{2pt}

\begin{tabularx}{\textwidth}{>{\bfseries}p{0.2\textwidth} >{\raggedright\arraybackslash}X}
\toprule
Phase & Criterion/Justification \\
\midrule

\multirow{2}{*}{\shortstack[l]{Phase One:\\ Core Concept \&\\ Character Drive}}
& \textbf{Attractiveness and Depth of the Premise:} The Premise is the conception that inspires the author to create a story. It is typically an open-ended question, such as “What would happen if…?”. A compelling premise can open the floodgates of imagination, making anything possible \textit{(pp.238)}. \\
\cmidrule{2-2}
& \textbf{Authenticity and Multi-dimensionality of Characters:} A true character is revealed in the choices they make under pressure; the greater the pressure, the deeper the revelation and the more the choice reveals their essential nature. A character’s dimensions and contradictions in their nature or behavior attract an audience's attention. Additionally, a protagonist may not only have a clear conscious desire, but also a self-contradictory subconscious desire, and these desires are the key drivers of the story \textit{(pp.211, 301, 379, 388)}. \\
\midrule

\multirow{2}{*}{\shortstack[l]{Phase Two:\\ World Setting \&\\ Inciting Incident}}
& \textbf{Preliminary Direction of the Controlling Idea:} The Controlling Idea is the ultimate meaning of the story, expressed through the action and aesthetic emotion of the climax. It should be a single, clear, coherent sentence that describes how and why life moves from its initial state to its final state. The controlling idea can shape the author's strategic choices, guiding what is appropriate or inappropriate during the creative process \textit{(pp.238, 246-248)}. \\
\cmidrule{2-2}
& \textbf{Refinement and Originality of Narrative Details:} It's essential to present a world that is internally consistent and credible in its scope, depth, and detail. Descriptions should be vivid and present, using the most specific, active verbs and concrete nouns to portray what readers can see (or hear) on screen. Dialogue should be lean, directional, and purposeful yet still sound like everyday conversation. The story can be enhanced through an Image System, which is a strategy of thematic imagery that runs through a film in a repetitive, subtle manner to deepen the complexity of aesthetic emotion on a subconscious level (pp.369, 389, 395-396, 401). \\
\midrule

\multirow{3}{*}{\shortstack[l]{Phase Three:\\ Structural Development \&\\ Conflict Escalation}}
& \textbf{Progressive Escalation of Conflict:} The story's structural function is to provide progressively increasing pressure, forcing characters to make increasingly difficult, risky choices and actions. This progressive complication means generating more and more conflict, pushing characters past the point of no return \textit{(pp.221-222, 405)}. \\
\cmidrule{2-2}
& \textbf{Breadth and Levels of Conflict:} A character’s world can be imagined as a series of concentric circles, which mark the levels of conflict in their life. Conflict can occur on three levels: internal (thoughts, body, emotions), interpersonal (family, friends, lovers), or extra-personal (social institutions, natural forces) . To achieve story complexity, the author should have the characters confront conflict on all three levels, often simultaneously \textit{(pp.215, 314-316)}. \\
\cmidrule{2-2}
& \textbf{Effectiveness of Structural Turning Points and Dynamic Rhythm:} A Scene is a micro-story that changes the charge of a value in a character’s life within a unity of time and space, through conflictual action . A Sequence is a series of scenes (usually two to five) whose climactic impact is greater than any preceding scene. An Act, in turn, is a series of sequences that turn on a major reversal in the value-charged state of a character's life. A well-told story will accelerate its pace, creating the rhythm of a life flow through a dynamic alternation of tension and release \textit{(pp.67, 74, 233, 290, 293)}. \\
\midrule

\multirow{3}{*}{\shortstack[l]{Phase Four:\\ Crisis, Climax \&\\ Controlling Idea\\ Confirmation}}
& \textbf{Dilemma of Crisis Decision and Character Consistency:} The Crisis, the third part of the five-part structure, signifies a decision. It is the moment of the protagonist's ultimate decision, where their next action will be the last and there will be no second chance. The crisis must be a true dilemma, requiring a choice between irreconcilable goods, or between the lesser of two evils, or a combination of both, which places the protagonist under the greatest pressure of their lives. The choice a protagonist makes in this dilemma will powerfully express their humanity and the world they inhabit, and this choice must be deeply consistent with the character's true nature \textit{(pp.211, 467, 599-600)}. \\
\cmidrule{2-2}
& \textbf{Impact and Depth of Meaning in the Climax:} The Story Climax, the fourth part of the five-part structure, is a climactic reversal that need not be filled with sound and fury, but must be filled with meaning. This means a total change of value, from positive to negative or negative to positive, with or without irony -- a value swing of maximum magnitude that is absolute and irreversible. A good climax should combine spectacle with truth, by focusing all meaning and emotion through a critical visual image \textit{(pp.570, 573-574)}. \\
\cmidrule{2-2}
& \textbf{Final Confirmation and Expression of the Controlling Idea:} The Controlling Idea is the ultimate meaning of the story, expressed through the action and aesthetic emotion of the climax. It is composed of two parts: Value plus Cause, which identifies the polarity of the story's key value at its climax, and points out the primary cause for the change in that value's final state. The final, dramatic action in the climax becomes the Controlling Idea of Value plus Cause, the purest expression of the story’s final and decisive meaning. The author must probe life to reveal new insights, new distillations of value and meaning. Only when the author profoundly believes in the meaning of their story can it be clearly and fully conveyed to the audience \textit{(pp.24, 238, 245-249, 261, 268, 293).} \\
\bottomrule
\end{tabularx}
\end{table*}

\section{Participant of Human Raters for Story Writing Evaluation}

Human ratings complemented self-reports. For story writing, we recruited 20 human raters (aged 20-30, \emph{M}=24.05, \emph{SD}=2.66) to assess story generation quality. Their demographic information is summarized in Table \ref{tab:demographic_human_eva}. These raters’ profiles are used solely to contextualize the human evaluation process.

\begin{table}[htb!]
\caption{Demographics of Human Raters for Story Writing Evaluation}
\label{tab:demographic_human_eva}
\centering
\begin{tabular}{@{}llll@{}}
\toprule
\textbf{ID} & \textbf{Gender} & \textbf{Age} & \textbf{AI Usage Frequency} \\ \midrule
P-1  & M & 20 & Heavy reliance \\
P-2  & M & 20 & Frequent \\
P-3  & M & 20 & Very frequent \\
P-4  & M & 23 & Frequent \\
P-5  & M & 23 & Very frequent \\
P-6  & F & 24 & Very frequent \\
P-7  & M & 23 & Very frequent \\
P-8  & F & 26 & Very frequent \\
P-9  & F & 26 & Frequent \\
P-10 & F & 24 & Frequent \\
P-11 & M & 24 & Very frequent \\
P-12 & F & 30 & Sometimes \\
P-13 & M & 25 & Sometimes \\
P-14 & M & 22 & Frequent \\
P-15 & F & 28 & Heavy reliance \\
P-16 & M & 23 & Frequent \\
P-17 & F & 24 & Very frequent \\
P-18 & F & 23 & Frequent \\
P-19 & M & 25 & Frequent \\
P-20 & F & 28 & Very frequent \\ \bottomrule
\end{tabular}
\end{table}

\section{System User Interface}

\begin{figure*}[!htbp]
  \centering
  \includegraphics[width=\linewidth]{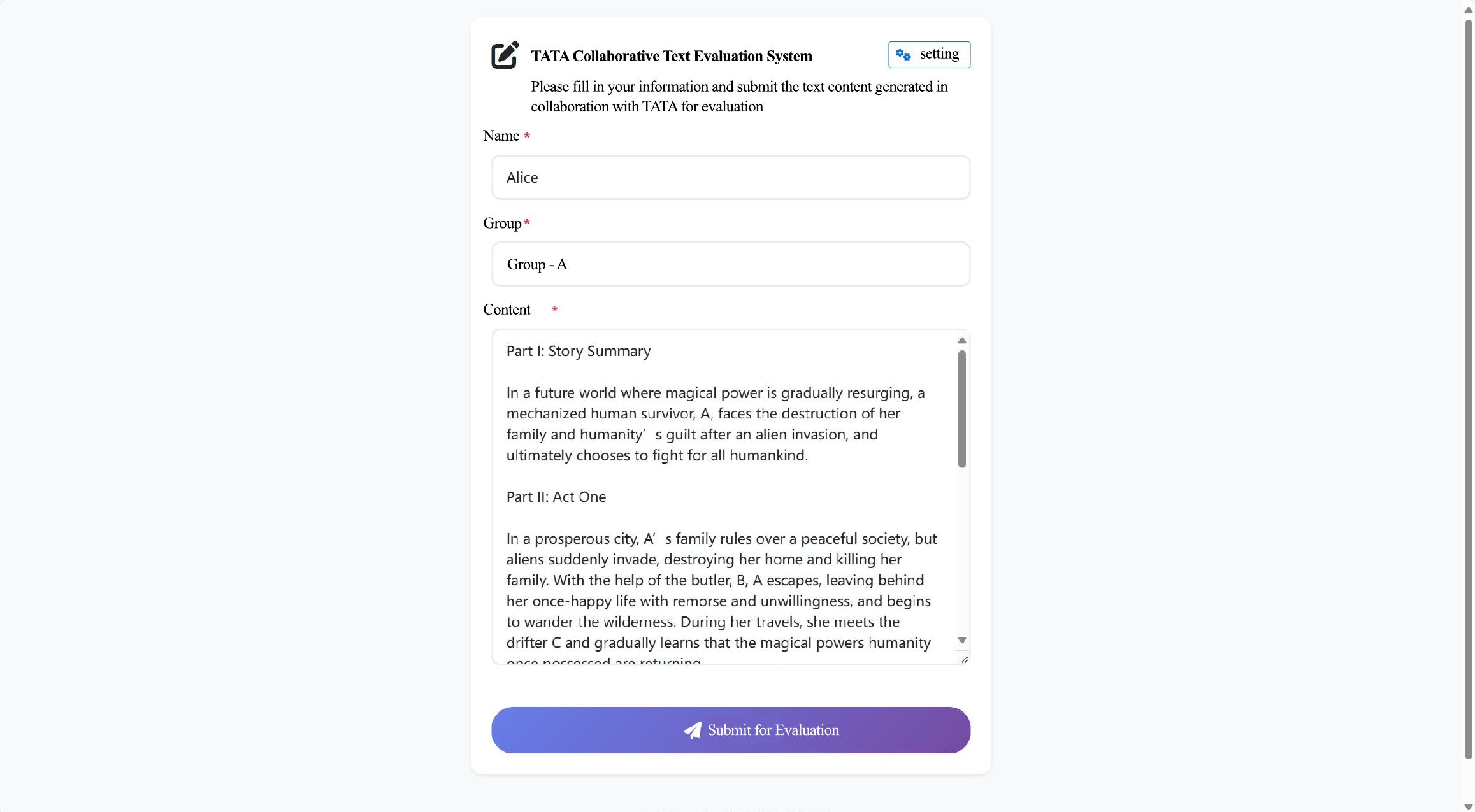}
  \caption{AI Story Co-Creation Dialogue}
 \label{fig:TATA AI Story Co-Creation Dialogue}
\end{figure*}

Figure \ref{fig:TATA AI Story Co-Creation Dialogue} shows the user interface where participants interact with the AI system, TATA, for story co-creation. The AI guides users through the multi-stage story generation process, provides prompts, and responds to user inputs in real time.

\begin{figure*}[!htbp]
  \centering
  \includegraphics[width=\linewidth]{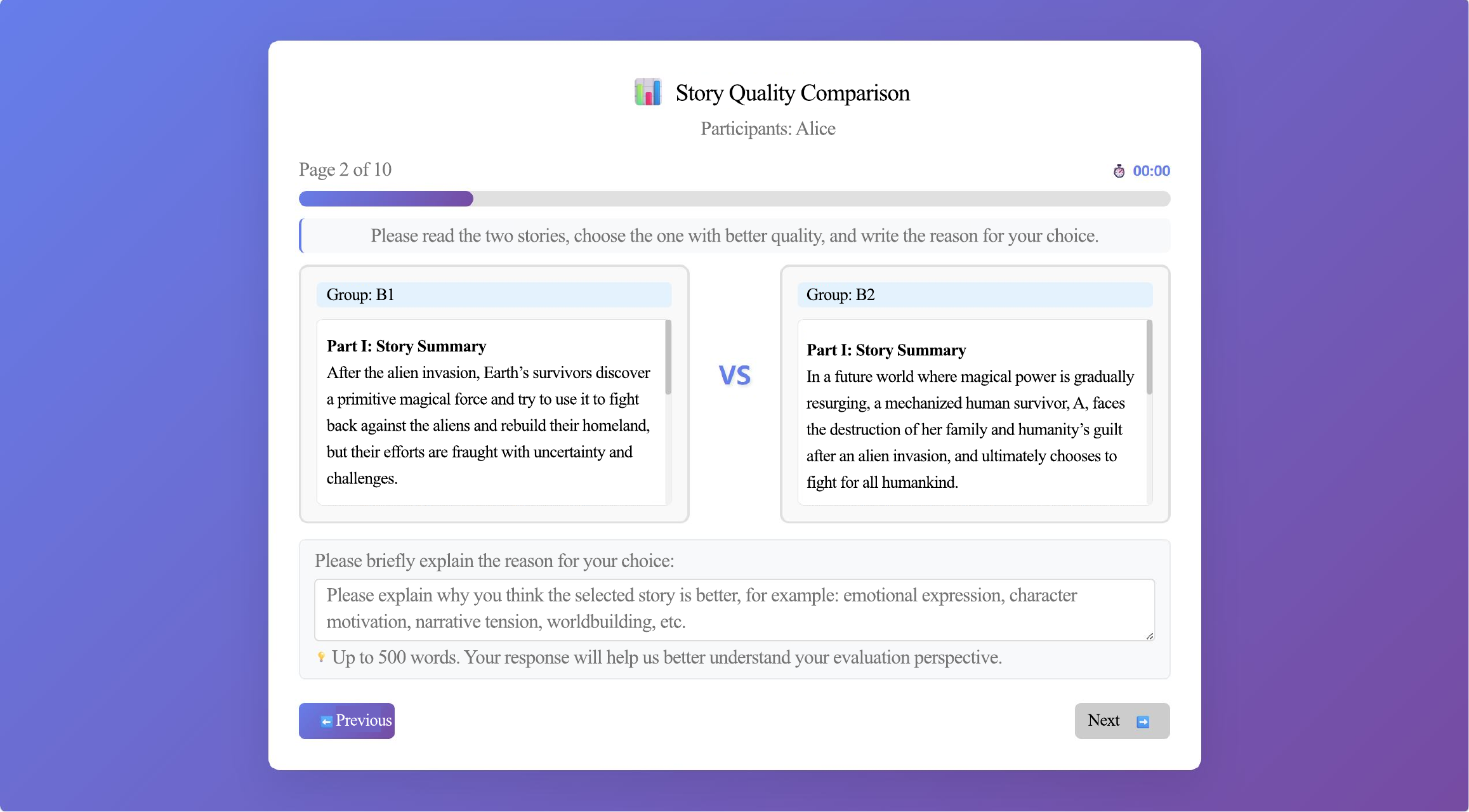}
  \caption{Story Evaluation Submission Form}
 \label{fig:Story Evaluation Submission Form}
\end{figure*}

Figure \ref{fig:Story Evaluation Submission Form} presents the submission interface where participants paste and submit the full story texts produced in collaboration with TATA for evaluation.

\begin{figure*}[!htbp]
  \centering
  \includegraphics[width=\linewidth]{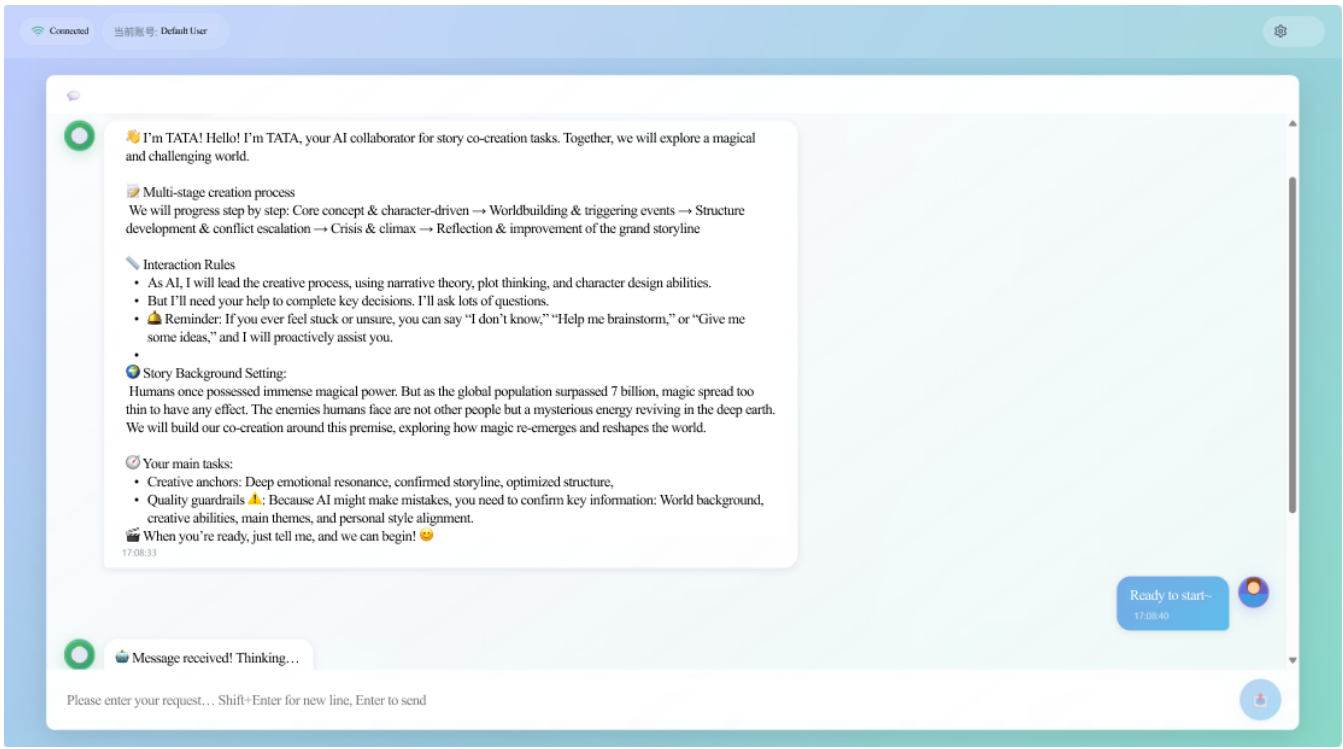}
  \caption{Story Quality Comparison Interface}
 \label{fig:Story Quality Comparison Interface}
\end{figure*}

Figure \ref{fig:Story Quality Comparison Interface} illustrates the evaluation interface where human raters compare pairs of story outputs, select the higher-quality story, and provide written justifications for their choice.

\section{Baseline System}
Participants interacted directly with a conventional tool-augmented LLM agent via the chat interface. As in typical LLM applications, every turn resulted in a model message returned to the participant, and all clarification and iteration occurred through this human-LLM dialogue. The backend model (GPT-4o), decoding parameters (including temperature), interface layout, memory/context handling, tool router, error/timeout policies, turn/time limits, and logging were identical to the Human Tool condition. The only manipulated factor was the ability to invoke humans as callable resources.

Domain tools matched the task: for travel planning, the Travel Planning Tool was enabled; for story writing, the Outline Generation Tool was enabled. No additional tools were added or removed beyond disabling any mechanism for human invocation.

Importantly, “calling a human as a tool” was not available in the baseline. The Human Tool framework components, “How to Define Human Tool,” “When to Call Human Tool,” and “How to Communicate with Human Tool”, were removed from the system prompt and tool stack. There were no UI affordances or backend routes to dispatch a subtask to a human or wait for a human reply mid-turn. If the model textually attempted such a request, it had no effect beyond appearing as text.

Behavior under uncertainty followed standard LLM practice: the agent asked the participant clarifying questions or proceeded with explicitly labeled, conservative assumptions. It did not solicit or defer to external human experts, reviewers, or collaborators (e.g., “ask a human editor/guide/evaluator”).

Everything else remained the same across conditions: GPT-4o version and parameters, interface elements, memory mechanism, tool permissions and retry logic, and evaluation/logging procedures. In short, the baseline is a normal human-LLM chat with domain tools; it excludes only the ability to route subtasks to separate human resources.

\label{sec:baseline}

\end{document}